\documentclass[onecolumn,a4paper,fleqn,usenatbib]{mnras}

\usepackage{newtxtext,newtxmath}
\usepackage[T1]{fontenc}
\usepackage{ae,aecompl}
\usepackage{upgreek}

\usepackage{amsmath}
\usepackage{mathtools}
\usepackage{graphicx}
\usepackage{float}
\usepackage{amsfonts}
\usepackage{amssymb}
\usepackage{makeidx}
\usepackage{mathrsfs}
\usepackage{subcaption}
\usepackage{natbib}
\usepackage{multirow}
\usepackage[normalem]{ulem}

\newcommand{\rs}{r_{_{\rm S}}}
\newcommand{\rL}{r_{_{\rm L}}}
\newcommand{\agsq}{a_{\rm th}^2}
\newcommand{\arsq}{a_{\rm rel}^2}

\newcommand{\gammag}{\gamma_{\rm th}}
\newcommand{\gammar}{\gamma_{\rm rel}}

\newcommand{\OmegaK}{\Omega_{\rm K}}

\newcommand{\msun}{M_\odot}

\newcommand\greens{f_{\rm G}}

\newcommand{\vel}{\upsilon}
\newcommand{\Eproton}{E_p}
\newcommand{\sigmaT}{\sigma_{_{\rm T}}}
\newcommand{\tauT}{\tau_{_{\rm T}}}
\newcommand{\yCM}{y_{_{\rm CM}}}

\usepackage{color}

\title[A Two-Fluid Model for Black-Hole Accretion Flows]{A Two-Fluid Model for Black-Hole Accretion Flows: Particle Acceleration, Outflows, and TeV Emission}

\author[J. Lee and P. A. Becker]{Jason P. Lee,$^1$\thanks{Email: je@gmu.edu}, Peter A. Becker,$^1$\thanks{Email: pbecker@gmu.edu}
\\
$^1$Department of Physics \& Astronomy,
George Mason University, Fairfax, VA 22030-4444, USA
}

\date{Accepted . Received ; in original form }

\pubyear{2019}

\begin{document}
\label{firstpage}
\pagerange{\pageref{firstpage}--\pageref{lastpage}}
\maketitle

\begin{abstract}
The multi-wavelength spectrum observed from M87 extends from radio wavelengths up to TeV $\gamma$-ray energies. The radio through GeV components have been interpreted successfully using SSC models based on misaligned blazar jets, but the origin of the intense TeV emission detected during flares in 2004, 2005, and 2010 remains puzzling. It has been previously suggested that the TeV flares are produced when a relativistic proton jet originating in the core of M87 collides with a molecular cloud (or stellar atmosphere) located less than one parsec from the central black hole. We explore this scenario in detail here using a self-consistent model for the acceleration of relativistic protons in a shocked, two-fluid ADAF accretion disc. The relativistic protons accelerated in the disc escape to power the observed jet outflows. The distribution function for the jet protons is used to compute the TeV emission produced when the jet collides with a cloud or stellar atmosphere. The simulated broadband radiation spectrum includes radio, X-ray and GeV components generated via synchrotron, as well as TeV emission generated via the production and decay of muons, positrons and electrons. The self-consistency of the model is verified by computing the relativistic particle pressure using the distribution function, and comparing it with the relativistic particle pressure obtained from the hydrodynamical model. We demonstrate that the model is able to reproduce the multi-wavelength spectrum from M87 observed by {\it VERITAS} and {\it HESS} during the high-energy flares in 2004, 2005, and 2010.
\end{abstract}




\section{INTRODUCTION}
\label{secIntro}
Active galactic nuclei (AGNs) and quasars are often observed to possess strong relativistic outflows, which are thought to be powered by accretion discs around supermassive black holes ($M\gtrsim10^8\,\msun$). Sources containing advection-dominated accretions flows (ADAFs) tend to produce strong radio and $\gamma$-ray emission, while radiating less efficiently in the X-ray region (e.g. Narayan et al. 1997; Yi \& Boughn 1998). On the other hand, sources with luminous X-ray emission tend to have weaker outflows (e.g. Yi \& Boughn 1999). ADAF discs occur when the accretion rate is far below the Eddington value, which inhibits efficient cooling, leading to gas temperatures approaching the virial value (e.g. Yi \& Boughn 1998). In this situation, the plasma is collisionless, meaning that the proton energy distribution is mediated by interactions with MHD waves (see Le \& Becker 2004, hereafter LB04; Le \& Becker 2005, hereafter LB05). 

In Lee \& Becker (2017, hereafter Paper~1), we explored the hydrodynamic, thermodynamic, and radiative properties of ADAF discs and the associated relativistic outflows. In this paper, we examine the implications of the disc and the outflows for the production of TeV $\gamma$-radiation, resulting from collisions between relativistic outflows and clouds or stelar atmospheres located within one parsec of the central black hole. Our specific focus here is on modeling and interpreting a series of high-energy flares observed from M87 by {\it VERITAS} and {\it HESS} in 2004, 2005, and 2010. The general scenario we consider has already been investigated by Barkov et al. (2012), based on studies of secondary emission developed by Kelner et al. (2006). However, these authors simply assumed the existence of a jet without postulating the physical mechanism for its formation. Here, we construct (for the first time as far as we can tell) a comprehensive, self-consistent model for the structure of the ADAF disc in M87, and explain how this disc powers a jet of relativistic protons, with such properties that the TeV emission observed in the flares can be understood as a consequence of the jet-cloud collision process studied by Barkov et al. (2012).

\subsection{Standing Shocks and Outflows in ADAF Discs}
\label{secStandShockADAF}
The possible existence of standing shocks and associated outflows in ADAF discs is a subject of ongoing debate. The question has been explored using a variety of steady-state and time-dependent simulations. For example, a number of previous studies have demonstrated that shocks can exist in both viscid and inviscid steady-state discs (e.g. Becker et al. 2011; Chakrabarti 1989; Chakrabarti \& Molteni 1993; Chattopadhyay \& Kumar 2016; Das, Becker \& Le 2009; Das, Chattopadhyay \& Chakrabarti 2001; Le \& Becker 2005; Lu \& Yuan 1997). Generally speaking, even when shocks can exist, there is usually a globally-smooth alternative solution. However, Becker \& Kazanas (2001) argued that when shocked and smooth solutions are both available, the second law of thermodynamics favors the formation of shocks because they tend to increase the total entropy of the system. In addition to the steady-state studies cited above, the possible existence of standing shocks in discs has also been explored using a series of  relativistic, time-dependent simulations. For example, Hawley, Smarr \& Wilson (1984a,b) and Chattopadhyay \& Kumar (2016) demonstrated that shocks do tend to form in hot tenuous discs. Okuda \& D. Molteni (2012) found that shocks may be unstable in their simulations of the accretion flow and outflow around Sgr~A*, but a subsequent detailed analysis by Le et al. (2016) focused on the stability of standing shocks in ADAF discs found that a stable mode exists over a broad range of the parameter space of viscosity and angular momentum. The underlying reason for the disagreement between these two sets of results is unclear, but it may reflect a difference in the choice of parameter values. For more recent similar studies supporting the existence of shocks in ADAF discs, see Dihingia et al. (2019), Kumar \& Gu (2019a,b), and Sarkar \& Chattopadhyay (2019). We also note that in the model under consideration here, shocks can exist over a broad range of the parameter space of angular momentum and energy of the accreted gas. The region of the parameter space within which shocks can form is indicated by the cream-colored region on the right side of Figure 5 in Paper~1.

The formation of a standing shock located near the centrifugal barrier in a collisionless ADAF disc creates an environment favorable for the acceleration of relativistic particles. Protons from the tail of the thermal ion distribution in the disc are able to cross the shock multiple times in the collisionless regime, leading to the formation of a population of nonthermal, relativistic particles via the first-order Fermi mechanism. Some of the accelerated protons escape from the disc in the vicinity of the shock, forming the observed jets, and removing binding energy from the disc, hence allowing accretion to proceed. Furthermore, it has been established that the acceleration of particles at a standing shock in the disc can be sufficient to power the observed strong outflows in radio-loud active galaxies containing supermassive black holes, such as M87, and also in the Galactic Centre source Sgr A* (Le \& Becker 2005; Becker et al. 2011). An alternative mechanism for the formation of jets and outflows from AGNs is provided by the Blandford-Znajek (1977) or Blandford-Payne (1982) mechanisms, which extract rotational energy from the black hole or accretion disc and convert it into a flux of electromagnetic fields and particles. However, recent studies have shown that this mechanism tends to channel energy into the equatorial plane of the disc, rather than along the rotation axis of the disc, calling into question of the effectiveness of these mechanisms for powering relativistic jets (e.g. Menon \& Dermer 2005; Le et al. 2018). This has stimulated renewed interest in the possibility of shock-driven outflows. In particular, we note that Le et al. (2016), Le et al. (2018), and Chattopadhyay \& Kumar (2016) confirmed the existence of stable disc/shock configurations with associated shock-driven outflows in ADAF discs.

\subsection{Multi-wavelength Observations of M87}
\label{secIntro1}
M87 has been observed for decades using a variety of instruments, covering emission generated in the radio, infrared, optical, and X-ray wavelengths (e.g. de Gasperin et al. 2012; Benkhali et al. 2019; Dermer \& Rephaeli 1988). Generally, the radio emission is concentrated in the central region and the halo of M87, the optical emission is dominated by starlight, and the hard X-ray emission originates in the core and the jet. Soft X-ray emission is observed from the halo of M87, probably produced via a combination of synchrotron emission and Compton up-scattering of disc photons (Dermer \& Rephaeli 1988). The GeV $\gamma$-ray emission form M87 is likely produced in the relativistic jet, and the TeV emission is probably produced in either the jet, or as a result of collisions between the jet and a cloud or stellar atmosphere (Benkhali et al. 2019). Abdo et al. (2009, hereafter A09) computed the comprehensive spectral energy distribution (SED) for M87 using a one-zone synchrotron self-Compton (SSC), covering emission from radio to GeV energies. The SED exhibits a double-hump structure, with a minimum flux at $\sim4\,$keV. In the standard SSC model, the low-frequency hump is attributed to synchrotron radiation, and the high-frequency feature results from the inverse-Compton scattering of a combination of externally produced radiation (from the disc or broad-line clouds), and internally produced synchrotron radiation (e.g. Finke et al. 2008). The {\it Fermi}-{\rm LAT} observed GeV emission from M87 during the 10 month all-sky survey running from 2008 August 4 - 2009 May 31 (A09). The intensity of the GeV emission observed by {\it Fermi}-{\rm LAT} varied by a factor of $\sim5$ over a timescale of $\sim$~two weeks.

Our main focus here is on the formation of the GeV-TeV $\gamma$-ray emission from M87 observed by {\it VERITAS} and {\it HESS}, which is not well explained by the standard SSC model. The TeV $\gamma$-ray emission from M87 exhibits both low and high luminosity states. Strong $\gamma$-ray emission in the TeV energy range was observed by {\it HESS} (Aharonian et al. 2006) when M87 was in a low state (2004) and a high state (2005). A very intense TeV flare was observed by {\it VERITAS} in 2010, during which the TeV flux was about an order of magnitude higher than that observed during the low state in 2004 (Aliu et al. 2012). It is important to emphasize that the GeV emission detected by {\it Fermi}-LAT was not contemporaneous with any of the TeV data sets obtained in 2004, 2005 and 2010. However, in the absence of any other GeV spectra, we will follow the example of A09 and Fraija \& Marinelli (2016) and use the {\it Fermi}-LAT data to constrain the multi-wavelength fits presented in Section \ref{secAstroApps}. We discuss the variable TeV emission in more detail below.

\subsection{TeV $\gamma$-Ray Flares}

From 2010 April 5-11, {\it VERITAS} observed a strong TeV flare from M87 with a duration of several days (Aliu et al. 2012). Two other TeV flares, with similar timescales, were observed by {\it HESS} in 2004 and 2005. These TeV flares present difficulties for the standard one-zone SSC model for blazar emission (Finke et al. 2008), which has motivated attempts to extend the standard model. For example, Lenain et al. (2008) proposed a scenario in which multiple plasma blobs containing highly relativistic electrons with Lorentz factor $\gamma\sim10^6$ propagate in the blazar jet with bulk Lorentz factor $\Gamma\sim10$. This type of model can account for the level of TeV emission observed by {\it HESS} in 2004 and 2005, although it requires a magnetic field $B\sim0.01-0.5\,$G, which is much lower than expected in the inner region of the jet (e.g. Vincent 2014; Sahu \& Palacios 2015). Alternatively, Tavecchio \& Ghisellini (2008) have explored a model in which the jet has two spatial components, with a relatively slow moving core producing the low-energy emission, and a very fast moving outer layer with bulk Lorentz factor $\Gamma\sim10^6$ producing the TeV emission. Although this model is capable of producing spectra that are comparable to those observed 2004 and 2005, there is no independent evidence for such a large bulk Lorentz factor in the M87 jet.

There are also models for the TeV emission that focus on hadronic processes (Benkhali et al. 2019). For example, Barkov et al. (2012, hereafter B12) argued that the {\it VERITAS} flare from M87 was the result of an interaction between a hadronic jet and a dense cloud, which could be the atmosphere of a red giant star, located $\sim0.01-0.1\,$pc from the black hole. Hadronic models present an attractive alternative to leptonic models, since hadrons are less strongly affected by synchrotron losses, which make it difficult to keep electrons sufficiently energized to produce TeV emission far from the central black hole. In the model of B12, relativistic protons power the observed TeV emission via proton-proton (pp) interactions, based on the nuclear physics formalism developed by Kelner et al. (2006, hereafter K06). Although B12 were able to roughly fit the 2010 {\it VERITAS} observations, they did not attempt to reproduce any other portion of the SED below an energy of $\sim 0.3\,$TeV. More importantly, B12 utilized an ad-hoc model for the proton distribution in the jet, which makes no connection with any acceleration mechanism or with the underlying accretion disc surrounding the supermassive black hole. A similar model was proposed by Fraija \& Marinelli (2016), in which the TeV $\gamma$-ray radiation results from pion production due to proton-photon (p$\gamma$) interactions. They used their model to fit the {\it HESS} 2004 data, but it was not applied to the interpretation of the {\it HESS} 2005 or {\it VERITAS} 2010 data. Moreover, Fraija \& Marinelli (2016) did not attempt to account for the production or acceleration of the inferred jet.

This situation has motivated us to examine the possibility that the two-fluid disc investigated in Paper~1 could be the source of the relativistic jet required to explain the production of the TeV emission observed from M87. Specifically, our goal is to determine whether the TeV $\gamma$-ray spectra observed in 2004, 2005, and 2010 can be explained as a natural consequence of a collision between a jet of relativistic protons (emanating from the two-fluid accretion disc) and a cloud or stellar atmosphere. If successful, the result would be a comprehensive, self-consistent model for the entire process, starting with the structure of the underlying accretion disc, and extending to the calculation of the properties of the jet outflow and the resulting TeV emission produced when the proton jet encounters the cloud. The organization of the paper is as follows. In Section \ref{secTwoFluidAccretion} we briefly review the accretion dynamics for our two-fluid model, and in Section~\ref{secSteadyStatePA} we solve the particle transport equation derived by LB07 to determine the proton distribution in the accretion disc in the context of our two-fluid model. In Section~\ref{secPartDist}, we present detailed applications using parameters appropriate for modeling the disc/outflows in M87. The production of secondary pions and $\gamma$-rays due to proton-proton collisions is analyzed in Section~\ref{secSecondaryRadiation}, and the model is applied to interpret the multi-wavelength emission from M87 in Section~\ref{secAstroApps}. Finally, in Section~\ref{secConclusion} we summarize our conclusions and discuss the astrophysical significance of our results.

\section{TWO-FLUID ACCRETION DYNAMICS}
\label{secTwoFluidAccretion}

\begin{figure}
\includegraphics[width=4in]{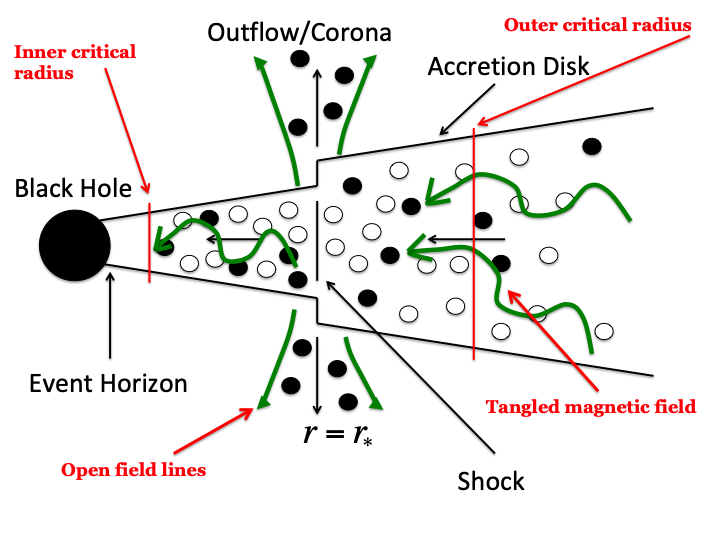}
\centering
\caption{Schematic diagram of our two-fluid disc/shock/outflow model, developed in Paper~1. The filled and open circles represent the accelerated protons and the MHD scattering centers, respectively. The scattering centers advect towards the black hole with the background flow velocity. Seed particles injected from the thermal population at the shock location are accelerated by crossing the shock multiple times. Protons escape from the disc into the corona at the shock location, forming a jet of relativistic plasma.}
\label{figMod3}
\end{figure}

In the scenario considered here, which was analyzed in detail in Paper~1, the plasma is gravitationally accelerated toward the central mass and encounters a standing shock just outside the centrifugal barrier (see Figure~\ref{figMod3}). The shock  is located in the supersonic region between the inner and outer critical points. The subsonic flow on the downstream side of the shock becomes supersonic again after passing through the inner critical point. Relativistic particles experience first-order Fermi acceleration in the vicinity of the shock discontinuity, producing a characteristic nonthermal distribution. A significant fraction of the accelerated particles are advected into the black hole, and the remainder escape by either diffusing in the outward radial direction through the disc, or by diffusing in the vertical direction to escape through the upper and lower surfaces of the disc near the shock radius.

In this section, we review some of the main features of the inviscid two-fluid model governing the accretion disc structure (see Paper~1 for complete details). The total energy transport rate in the disc in the inward radial direction, $\dot E$, is defined by
\begin{equation}
\dot E = \dot E_{\rm th} + \dot E_{\rm rel} \ , 
\end{equation}
where the energy transport rates for the thermal and relativistic particle populations are given, respectively, by
\begin{equation}
\dot E_{\rm th} = \dot M \left(\frac{1}{2} \vel^2 + \frac{1}{2}\frac{\ell_0^2}{r^2} + \Phi + \frac{\agsq}{\gamma_{\rm th} - 1}\right) \ , 
\label{eqDeltaEG}
\end{equation}
and
\begin{equation}
\dot E_{\rm rel} = \dot M \left(\frac{\arsq}{\gamma_{\rm rel} - 1} + \frac{\kappa}{\rho\vel}\frac{dU_{\rm rel}}{dr}\right) \ .
\label{eqDeltaEr}
\end{equation}
Here, $\dot M$ represents the accretion rate onto the black hole, $a_{\rm th}$ and $a_{\rm rel}$ denote the adiabatic sound speeds for the thermal and relativistic particles, respectively, $\vel > 0$ is the radial inflow speed, $\rho$ is the mass density, $\ell_0$ is the angular momentum per unit mass, $\kappa$ is the radial diffusion coefficient for the relativistic particles, $U_{\rm rel}$ is the relativistic particle energy density, $\Phi$ is the pseudo-Newtonian gravitational potential, and $\gamma_{\rm th}=5/3$ and $\gamma_{\rm rel}=4/3$ denote the adiabatic indices for the thermal and relativistic particles, respectively. The pseudo-Newtonian potential is defined by (Paczy\'nski \& Wiita 1980)
\begin{equation}
\Phi = - \frac{G M}{r - \rs} \ ,
\label{eqPhi}
\end{equation}
where $\rs = 2GM/c^2 = 2 R_g$ represents the Schwarzschild radius of the black hole and $R_g$ denotes the gravitational radius. In the two-fluid model, pressure support is provided by both the thermal gas and the relativistic particles, and therefore the disc half-thickness is given by (see Paper~1)
\begin{equation}
H(r) = \frac{1}{\OmegaK}\left(\frac{\gamma_{\rm th}}{\gamma_{\rm rel}}a^2_{\rm rel}+a^2_{\rm th}\right)^{1/2} \ .
\label{eqH}
\end{equation}

The radial diffusion coefficient $\kappa$ appearing in equation~(\ref{eqDeltaEr}) describes the scattering of relativistic particles by MHD waves, and is computed using (see equation~13 from Paper~1)
\begin{equation}
\kappa(r) = \kappa_0 \, \vel(r)\rs \left(\frac{r}{\rs}-1\right)^2 \ , 
\label{eqKappa}
\end{equation}
where $\kappa_0$ is a dimensionless constant. The total energy transport rate $\dot E$ is constant throughout the disc, except at the shock location, where there is a jump in $\dot E$ due to the escape of energy from the disc, represented by
\begin{equation}
L_{\rm jet} = -\Delta\dot E = -\frac{1}{2} \dot M \Delta\vel^2 \ ,
\label{eqDeltaE}
\end{equation}
where $L_{\rm jet}$ denotes the jet kinetic luminosity, and the operator $\Delta$ is defined by
\begin{equation}
\Delta[f] \equiv \lim_{\delta\to 0} f(r_*-\delta) - f(r_*+\delta) = f_+ - f_- \ ,
\label{eqDeltaM2}
\end{equation}
with the subscripts ``+'' and ``-'' denoting post-shock and pre-shock values, respectively, for any physical quantity.

As discussed in Section 3 from Le \& Becker (2005), an isothermal shock will produce a larger compression ratio than either a Rankine-Hugoniot shock (which conserves energy flux), or an isentropic shock (which conserves entropy). It follows that only the isothermal or isentropic shocks are capable of radiating the energy required to power an outflow. The detailed analysis carried out in Paper~1 indicates that the compression ratio in the application of our model to M87 is not large, $R\sim1.7$, and therefore an isentropic shock would probably yield very similar results to the isothermal shock case, which is assumed here. As the gas crosses the standing shock, the decrease in the relativistic particle sound speed (due to the escape of energy into the outflow) leads to the condition
\begin{equation}
a_{{\rm rel}+} <  a_{{\rm rel}-} \ .
\label{eqarel}
\end{equation}
On the other hand, since the shock is assumed to be isothermal, the jump condition for the thermal particle sound speed is given by
\begin{equation}
a_{{\rm th}+} = a_{{\rm th}-} \ .
\label{eqath}
\end{equation}
Consideration of equations (\ref{eqH}), (\ref{eqarel}), and (\ref{eqath}) leads to the conclusion that in the two-fluid model, there is a decrease in the disc half-thickness as the gas crosses the shock, as depicted schematically in Figure~\ref{figMod3}.

\section{PARTICLE ACCELERATION AND TRANSPORT EQUATION}
\label{secSteadyStatePA}

Our goal in this paper is to analyze the transport and acceleration of relativistic particles (protons) in a disc governed by the two-fluid dynamical model developed in Paper~1. The particle transport model for the relativistic protons in the disc includes terms describing spatial diffusion, advection, particle escape, and first-order Fermi energization. The solution to the transport equation is the steady-state Green's function, $\greens(\Eproton,r)$, describing the particle distribution in the disc resulting from monoenergetic particle injection, where $\Eproton$ and $r$ denote the proton energy and the radius in the disc, respectively. Our formalism is similar to the one employed by LB07 in the context of their one-fluid dynamical model. However, an important distinction is that we are including the effect of the relativistic particle pressure on the dynamical structure of the disc, which was neglected by LB07. Hence we will need to reexamine some of the fundamentals described in LB07 in order to create a self-consistent model, which is one of our primary objectives in this paper.

The issue of the magnetic topology at the base of the flow warrants further discussion. Various studies indicate that astrophysical outflows tend to occur along open field lines that are anchored in active regions. For example, de Gouveia Dal Pino et al. (2010) argued that outflows from accretion discs in AGNs occur along open field lines, and are powered by particle acceleration occurring in regions experiencing violent magnetic reconnection. These regions are likely to be concentrated in the vicinity of shocks because shocks tend to enhance the magnetic shear that leads to reconnection. An analogous process in the context of solar flares was suggested by Plotnikov et al. (2017), who argued that magnetic reconnection in the vicinity of coronal shocks both creates open magnetic field lines and also powers the strong $\gamma$-ray emission observed in some solar flares. Desai \& Burgess (2008) invoked a similar mechanism in their study of coronal mass ejection-driven particle acceleration at Earth's bow shock. In both the AGN and solar applications, the wind or jet outflow starts off at the base with a non-relativistic velocity, and then subsequently expands as it accelerates to a highly relativistic terminal velocity. This is further discussed in Section~\ref{subEscPart}.

The particle transport formalism used in this work follows the approach of LB07, which treats the particle distribution function $\greens$ as a vertical average over the disc half-thickness, denoted by $H(r)$. We assume that the isothermal shock radius, $r_*$, is also the location of the particle injection from the tail of the thermal background, triggered by magnetic reconnection in the vicinity of the shock (e.g. Drury 2012; Jones \& Ellison 1991). Following Desai \& Burgess (2008), we assume that the escape of the relativistic particles from the disc into the corona and outflow is also concentrated at the shock radius, due to the presence of open field lines in the vicinity of the velocity discontinuity, as indicated in Figure~\ref{figMod3}. This establishes a connection between the jump in the relativistic energy flux and the energy carried away by the escaping particles at the shock location, hence ensuring self-consistency between the dynamical model and the particle transport calculation.

\subsection{Transport Equation}
\label{subTE}

The Green's function, $\greens(\Eproton,r)$, describes the energy and spatial distribution of relativistic protons in the accretion disc, resulting from the continual injection of $\dot{N}_0$ seed particles per unit time with energy $E_0$ from a source located at radius $r_*$. The corresponding relativistic particle number and energy densities, $n_{\rm rel}(r)$ and $U_{\rm rel}(r)$, respectively, are related to $\greens(\Eproton,r)$ via the integrals
\begin{equation}
n_{\rm rel}(r) = \int_{E_0}^\infty 4\uppi \Eproton^2\,\greens(\Eproton,r)\,d\Eproton \ , \qquad
U_{\rm rel}(r) = \int_{E_0}^\infty 4\uppi \Eproton^3\,\greens(\Eproton,r)\,d\Eproton \ .
\label{eqNrUrGreens}
\end{equation}
The relativistic particle pressure $P_{\rm rel}$ is related to the energy density $U_{\rm rel}$ via $P_{\rm rel} = (\gamma_{\rm rel}-1) U_{\rm rel}$, where $\gamma_{\rm rel}=4/3$.

The lower bound for the integrations over the proton energy in equations~(\ref{eqNrUrGreens}) is set to $E_0$ because there is no deceleration included in the model considered here. The vertically-averaged form of the transport equation satisfied by the Green's function can be written as (see equation~B3 from Paper~1),
\begin{equation}
H\vel_r\frac{\partial\greens}{\partial r} = \frac{1}{3r}\frac{\partial}{\partial r}(r H \vel_r)\Eproton\frac{\partial\greens}{\partial \Eproton}
+ \frac{1}{r}\frac{\partial}{\partial r}\left(rH\kappa\frac{\partial\greens}{\partial r}\right)
+ \frac{\dot N_0\delta(\Eproton-E_0)\delta(r-r_*)}{(4\uppi\, E_0)^2 r_*} - \dot{f}_{\rm esc} \ , 
\label{eqVertTransport}
\end{equation}
where $\vel_r = - \vel < 0$ is the inflow velocity, $\kappa$ is the radial diffusion coefficient in the disc (describing the scattering of relativistic protons with MHD waves), $c$ is the speed of light, and the $\dot f_{\rm esc}$ is the escape term, defined by
\begin{equation}
\dot{f}_{\rm esc} = A_0 c H_*\delta(r-r_*)\greens \ .
\label{eqEscape}
\end{equation}
The dimensionless constant $A_0$ sets the efficiency of the escape of particles from the disc at the shock location, and is computed using the energy conservation relation $L_{\rm jet}=L_{\rm esc}$ (see equations~131 and 133 from Paper~1)
\begin{equation}
A_0 = \frac{L_{\rm jet}}{4\uppi r_*H_*cU_{\rm rel}(r_*)} \ ,
\end{equation} 
where $H_*=H(r_*)$ and $U_{\rm rel}(r_*)$ represent the disc half-thickness and the relativistic particle energy density, respectively, at the shock location. In the vicinity of the shock, the inflow speed $\vel = -\vel_r > 0$ is discontinuous, and is denoted by (see equation~23 from LB07)
\begin{equation}
\frac{d\vel}{dr}\to(\vel_--\vel_+)\delta(r-r_*) \ , \qquad r\to r_* \ , 
\label{eqDvDrDelta}
\end{equation}
where $\vel_-$ and $\vel_+$ represent the positive inflow speeds just upstream and downstream from the shock, respectively. Due to the velocity discontinuity, the first-order Fermi acceleration of the particles is concentrated in the region surrounding the shock.

\subsection{Separation Functions}
\label{subSpatialSep}

For proton energies $\Eproton > E_0$, the source term in equation~(\ref{eqVertTransport}) vanishes, and the resulting equation is homogeneous and separable in terms of the functions
\begin{equation}
f_\lambda(\Eproton,r) = \left(\frac{\Eproton}{E_0}\right)^{-\lambda} Y(r) \ ,
\label{eqSeparationFunction}
\end{equation}
where $\lambda$ is the separation constant, and the spatial separation functions $Y(r)$ satisfy the differential equation
\begin{equation}
- H\vel\frac{dY}{dr} = \frac{\lambda_n}{3r}\frac{d}{dr}(rH\vel)Y + \frac{1}{r}\frac{d}{dr}\left(rH\kappa\frac{dY_n}{dr}\right)
- A_0 c H_*\delta(r-r_*)Y \ .
\label{eqDYnDr2}
\end{equation}
Using equation~(\ref{eqKappa}) to substitute for $\kappa$ in equation~(\ref{eqDYnDr2}) yields
\begin{equation}
\frac{d^2 Y}{dr^2} + \left[\frac{\rs}{\kappa_0(r-\rs)^2} + \frac{d\ln(rH\vel)}{dr} + \frac{2}{r-\rs}\right]\frac{dY}{dr}
+ \frac{\lambda_n\rs Y}{3\kappa_0(r-\rs)^2}\frac{d\ln(rH\vel)}{dr} = 0 \ ,
\label{eqDYnDr}
\end{equation}
which is identical to equation~(30) from LB07 for their one-fluid model; hence it is also valid in the case of our two-fluid model. However, it should be emphasized that the dynamical profiles for $H(r)$ and $\vel(r)$ used here are significantly different from those adopted by LB07 in their one-fluid model, since our two-fluid model includes the effect of the relativistic particle pressure on the background flow.

\subsection{Jump Conditions}
\label{SecJump}

The global solutions for the spatial separation function $Y(r)$ must satisfy equation~(\ref{eqDYnDr}), in addition to a set of physical boundary and jump conditions. The jump conditions are associated with the existence of the shock/source at radius $r=r_*$, and can be obtained by integrating equation~(\ref{eqDYnDr}) with respect to radius $r$ in a small region surrounding the shock location. The results obtained are (see Appendix~\ref{AppendixEigenJump}) the continuity condition,
\begin{equation}
\Delta[Y] = 0 \ ,
\label{eqDeltaYn1}
\end{equation}
and the derivative jump condition,
\begin{equation}
\Delta\left[\frac{\lambda_n}{3}H\vel Y + H\kappa\frac{dY}{dr}\right] = - A_0 c H_* Y(r_*) \ ,
\label{eqDeltaYn2}
\end{equation}
where $\Delta$ represents the difference between post-shock and pre-shock quantities (see equation~\ref{eqDeltaM2}).

\subsection{Spatial Eigenfunctions}
\label{subSpatialEigen}

The spatial eigenfunctions, $Y_n(r)$, are those special instances of the separation function $Y(r)$ that satisfy the differential equation~(\ref{eqDYnDr}) as well as the physical boundary and jump conditions. The procedure required to obtain the global solution for the eigenfunction $Y_n(r)$ involves two separate integrations in the inner and outer regions, yielding two fundamental solutions, denoted by $G_n^{\rm in}(r)$ and $G_n^{\rm out}(r)$, respectively. The global solution for $Y_n(r)$ is then developed by combining the fundamental solutions $G_n^{\rm in}(r)$ and $G_n^{\rm out}(r)$, which yields
\begin{equation}
Y_n(r) = \begin{cases}
G_n^{\rm in}(r) \ ,  & r\leq r_* \ , \\ 
a_n G_n^{\rm out}(r) \ ,  & r\geq r_* \ , \\
\end{cases}
\label{eqYnCases}
\end{equation}
where the matching coefficient, $a_n$, is computed using
\begin{equation}
a_n = \frac{G_n^{\rm in}(r_*)}{G_n^{\rm out}(r_*)} \ ,
\label{eqAn}
\end{equation}
which ensures the continuity of $Y_n(r)$ at the shock location, $r=r_*$, as required by equation~(\ref{eqDeltaYn1}).

\subsection{Boundary Conditions and Eigenvalues}
\label{subBCs}

The spatial eigenfunctions $Y_n(r)$ must also satisfy a set of boundary conditions, which yields a discrete set of values for $\lambda$, denoted by the eigenvalues, $\lambda_n$. To develop the inner boundary condition, applicable close to the event horizon ($r \to \rs$), we note that near the horizon in the two-fluid model, the plasma behaves adiabatically, because diffusion becomes negligible as the flow velocity approaches $c$ (e.g. Paper~1; Weinberg 1972). Furthermore, the thermal particle pressure dominates over the relativistic particle pressure as $r \to \rs$, because $\gammag > \gammar$. Hence, we can adopt the asymptotic relation derived by LB07, which states that near the horizon, the behavior of $G_n^{\rm in}(r)$ is given by (see equation~35 from LB07),
\begin{equation}
G_n^{\rm in}(r)\to g_n^{\rm in}(r)\equiv\left(\frac{r}{\rs}-1\right)^{-\lambda_n/(3\gamma_{\rm th}+3)} \ , \qquad r\to\rs \ .
\label{eqGin}
\end{equation}
Likewise, the asymptotic form applicable at large radii ($r\to\infty$), where spatial diffusion dominates, is derived in Appendix \ref{AppendixEigenAsymptotic}. The result obtained is 
\begin{equation}
G_n^{\rm out}(r)\to g_n^{\rm out}(r)\equiv \frac{C_1}{r}+1 \ , \qquad r\to\infty \ , 
\label{eqGout1}
\end{equation}
where $C_1$ is a constant. Since the particle transport is dominated by diffusion at large radii, it follows that $G_n^{\rm out}(r)\propto U_{\rm rel}(r)$ as $r\to\infty$. Based on equation~(116) from Paper~1, the asymptotic behavior of $U_{\rm rel}(r)$ is therefore given by
\begin{equation}
U_{\rm rel}(r) \to U_{\rm rel,\infty}\left(\frac{C_1}{r}+1\right) \ , \qquad r \to \infty \ .
\label{eqUasympOut}
\end{equation}
The global numerical solution for $U_{\rm rel}(r)$ was already obtained as part of the set of hydrodynamical model results computed in Paper~1, and therefore the constant $C_1$ can be calculated using the hydrodynamical results. Incorporating the resulting value of $C_1$ into equation~(\ref{eqGout1}) allows us to compute the asymptotic behavior of $G_n^{\rm out}$, so that we obtain for the outer boundary condition
\begin{equation}
G_n^{\rm out}(r)\to g_n^{\rm out}(r) = \frac{U_{\rm rel}(r)}{U_{\rm rel,\infty}} \ , \qquad r\to\infty \ .
\label{eqGout}
\end{equation}
The validity of the asymptotic forms in equations~(\ref{eqGin}) and (\ref{eqGout}) is demonstrated in Appendix \ref{AppendixEigenAsymptotic} by comparing the numerical solutions obtained for the spatial eigenfunctions with the predicted asymptotic forms. The results are similar to those depicted in Figures~3 and 4 from LB07.

\subsection{Green's Function Solution}
\label{subGFS}

Once the inner and outer fundamental solutions, $G_n^{\rm in}(r)$ and $G_n^{\rm out}(r)$, respectively are determined via numerical integration of equation~(\ref{eqDYnDr}) in the inner and outer regions of the disc, the matching coefficient, $a_n$, is computed using equation~(\ref{eqAn}), and the general solution for the spatial eigenfunction $Y_n(r)$ is evaluated using equation~(\ref{eqYnCases}). In general, the boundary conditions and the jump conditions are not satisfied for arbitrary values of the separation constant $\lambda$. Hence, $\lambda$ must be varied in order to determine the discrete eigenvalues, denoted by $\lambda_n$. The process is repeated for integer values $n$, starting with $n=1$, until the desired number of eigenvalues and eigenfunctions is obtained. We verify the orthogonality of the spatial eigenfunctions $Y_n(r)$ computed using equation~(\ref{eqYnCases}) in Appendix \ref{AppendixOrtho}, and apply this technique to M87 in Section \ref{secAstroApps}. Since the eigenfunctions form an orthogonal set, it follows that we can develop a series expansion for the Green's function, $\greens(\Eproton,r)$, by writing
\begin{equation}
\greens(\Eproton,r) = \sum^{N_{\rm max}}_{n=1}b_nY_n(r)\left(\frac{\Eproton}{E_0}\right)^{-\lambda_n} \ , \quad \Eproton \geq E_0 \ , 
\label{eqGreensExpansionV1}
\end{equation}
where the expansion coefficients, $b_n$, are derived in Appendix~\ref{secAppExp} and computed using equation~(\ref{eqD10}). The Green's function represents the proton distribution in the disc resulting from the continual injection of seed protons with energy $E_0$ at radius $r_*$. In our model, the seed protons are injected from the tail of the thermal Maxwellian in the vicinity of the shock, or as the result of magnetic reconnection (e.g. Paper~1, LB05). In our application to M87, we generally set $N_{\rm max}=10$, which yields an accuracy of $\sim5\%$ based on the convergence properties of the expansion in equation~(\ref{eqGreensExpansionV1}).

\section{PARTICLE DISTRIBUTION IN THE M87 JET}
\label{secPartDist}

\begin{table}
\centering
\caption{Disc structure parameters, originally shown in Paper~1. All quantities are expressed in gravitational units $(GM=c=1)$.}
\begin{tabular}[width=1\columnwidth]{lccccccccrcccc}
\hline\hline
Model & $\ell_0$ & $\kappa_0$ & $\Delta\epsilon$ & $r_*$ & $H_*$ & $A_0$ & $\eta_{_{\rm S}}$ & $\eta$ & $K_{\rm th}/K_{\rm rel}$ & $K_{\rm th}$ & $K_{\rm rel}$\\
\hline
A & 3.1340 & 0.02044 & -0.005671 & 12.565 & 6.20 & 0.050 & 6.63 & 5.95 & 7,400 & $3.04\times10^{-3}$ & $4.10\times10^{-7}$\\
B & 3.1524 & 0.02819 & -0.005998 & 11.478 & 5.46 & 0.052 & 6.41 & 3.65 & 7,700 & $2.79\times10^{-3}$ & $3.63\times10^{-7}$\\
C & 3.1340 & 0.03000 & -0.006427 & 14.780 & 7.49 & 0.100 & 3.56 & 3.84 & 65,000 & $3.64\times10^{-3}$ & $5.61\times10^{-8}$\\
D & 3.1524 & 0.05500 & -0.006116 & 14.156 & 6.91 & 0.125 & 1.42 & 1.45 & 260,000 & $3.51\times10^{-3}$ & $1.35\times10^{-8}$\\
\hline
\end{tabular}
\label{table:table1} 
\end{table}
\begin{table}
\centering
\caption{Model Energy Parameters, originally shown in Paper~1.}
\begin{tabular}[width=1\columnwidth]{lccrcccccccccc}
\hline\hline
Model & $\Gamma_\infty$ & $L_{\rm jet}\,\left({\rm erg \ s}^{-1}\right)$ & $\dot N_0\,\left({\rm s}^{-1}\right)$ & $\dot N_{\rm I}\,\left({\rm s}^{-1}\right)$ & $\dot N_{\rm II}\,\left({\rm s}^{-1}\right)$ & $\dot N_{\rm esc}\,\left({\rm s}^{-1}\right)$ & $\dot M\,\left(\msun\,{\rm yr}^{-1}\right)$ & $\dot N_{\rm th}\,\left({\rm s}^{-1}\right)$\\
\hline
A & 3.52 & 5.5$\times10^{43}$ & $2.75 \times 10^{46}$ & $7.42\times10^{43}$ & $-1.61\times10^{46}$ & $1.04\times10^{46}$ & 1.71$\times10^{-1}$ & 6.46$\times10^{48}$\\
B & 3.52 & 5.5$\times10^{43}$ & $2.75 \times 10^{46}$ & $3.49\times10^{44}$ & $-1.57\times10^{46}$ & $1.04\times10^{46}$ & 1.62$\times10^{-1}$ & 6.11$\times10^{48}$\\
C & 2.46 & 5.5$\times10^{43}$ & $2.75 \times 10^{46}$ & $1.37\times10^{45}$ & $-9.62\times10^{45}$ & $1.49\times10^{46}$ & 1.51$\times10^{-1}$ & 5.70$\times10^{48}$\\
D & 2.48 & 5.5$\times10^{43}$ & $2.75 \times 10^{46}$ & $5.46\times10^{45}$ & $-6.71\times10^{45}$ & $1.48\times10^{46}$ & 1.59$\times10^{-1}$ & 5.99$\times10^{48}$\\
\hline
\end{tabular}
\label{table:table1a} 
\end{table}

In Paper~1, we investigated the acceleration of relativistic particles in an inviscid ADAF disc containing a standing, isothermal shock. The focus of that study was the determination of the self-consistent velocity distribution in the disc, including the back-reaction exerted on the flow by the pressure of the accelerated relativistic particles. We found that the inclusion of the particle pressure tends to create a smooth precursor deceleration region on the upstream side of the shock, similar to that seen in the structure of cosmic-ray modified shocks (e.g. Axford et al. 1977; Becker \& Kazanas 2001). For a given source with a measured jet kinetic power, $L_{\rm jet}$, and a known black hole mass, $M$, we found that for a specific value of the diffusion parameter, $\kappa_0$ (see equation~\ref{eqKappa}), several distinct flow solutions can be obtained for different values of the accreted entropy ratio, $K_{\rm th}/K_{\rm rel}$, which denotes the ratio of the thermal and relativistic particle entropy parameters at the event horizon. We focused on four particular models in Paper~1, based on the values for the specific angular momentum $\ell_0$ and the diffusion parameter $\kappa_0$ adopted in models 2 and 5 from LB05, combined with variation of $\kappa_0$ in order to maximize the terminal Lorentz factor, $\Gamma_{\infty}$, for the escaping particles. In Tables \ref{table:table1} and \ref{table:table1a}, we list all of the relevant parameter values for each of the four models examined in Paper~1. Note that the values reported in Table \ref{table:table1a} for $\Gamma_{\infty}$ are slightly different from those obtained in Paper~1, which is due to a slight improvement in the accuracy of the numerical algorithm used in the computations performed here.

It is important to discuss the physical significance of the entropy ratio $K_{\rm th}/K_{\rm rel}$. The entropy per particle for the thermal gas is given by (see equation 28 from Paper~1)
\begin{equation}
K_{\rm th} \equiv r^{3/2}(r-\rs) \vel a_{\rm th}^{2/(\gamma_{\rm th}-1)}\left(\frac{\gamma_{\rm th}}{\gamma_{\rm rel}} a^2_{\rm rel}
+ a^2_{\rm th}\right)^{1/2} \ .
\label{eqKg}
\end{equation}
Likewise, the entropy per particle for the relativistic population is given by (see equation 29 from Paper~1)
\begin{equation}
K_{\rm rel} \equiv r^{3/2}(r-\rs) \vel a_{\rm rel}^{2/(\gamma_{\rm rel}-1)}\left(\frac{\gamma_{\rm th}}{\gamma_{\rm rel}} a^2_{\rm rel}
+ a^2_{\rm th}\right)^{1/2} \ .
\label{eqKr}
\end{equation}
The values for $K_{\rm th}$ and $K_{\rm rel}$ at the inner boundary ($r=2.1\,R_g$) are reported in Table \ref{table:table1}. The total rate of accretion of entropy onto the black hole is given by the weighted sum $\dot N_{\rm th}K_{\rm th}+\dot N_{\rm rel}K_{\rm rel}$, where $\dot N_{\rm rel}$ and $\dot N_{\rm th}$ represent the accretion rates at the horizon for the relativistic particles and the thermal gas, respectively. The thermal particle accretion rate is related to the mass accretion rate $\dot M$ via $\dot N_{\rm th}=\dot M/m_p$, and the relativistic particle accretion rate is given by $\dot N_{\rm rel}=\left|\dot N_{\rm II}\right|$, where the absolute value is taken because $\dot N_{\rm II}$ is a negative quantity (see Table \ref{table:table1a}). Computing the total system entropy using the weighted sum indicates that the system entropy is dominated by the thermal particles. Furthermore, we find that the total entropy accretion rate is maximized for Model C, and therefore we will focus exclusively on Model C in this study.

\begin{figure}
\centering
\begin{subfigure}[b]{.5\textwidth}
\centering
\includegraphics[width=1\linewidth]{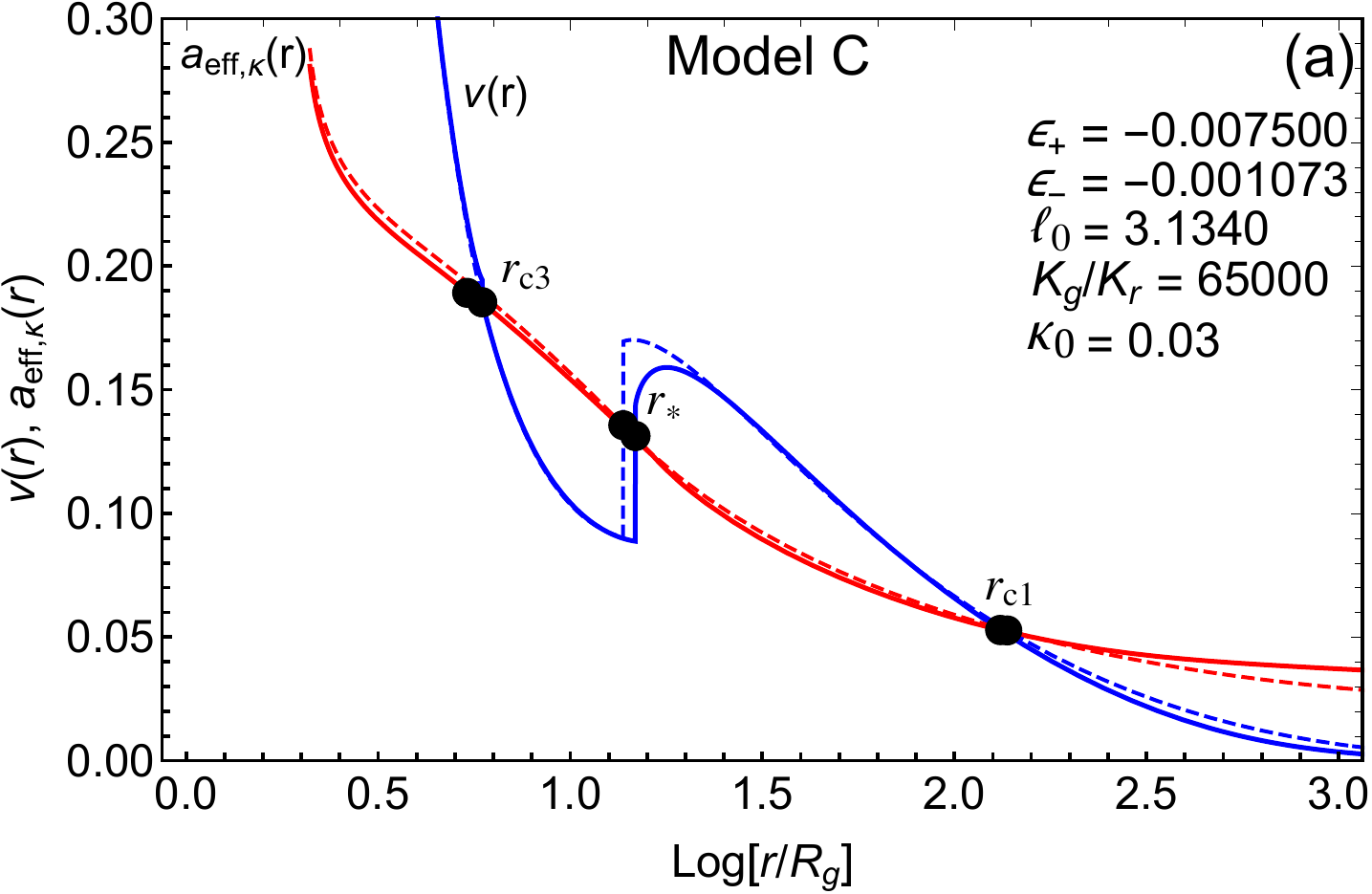}
\label{fig:1sub1}
\end{subfigure}
\quad
\begin{subfigure}[b]{.32\textwidth}
\includegraphics[width=1\linewidth]{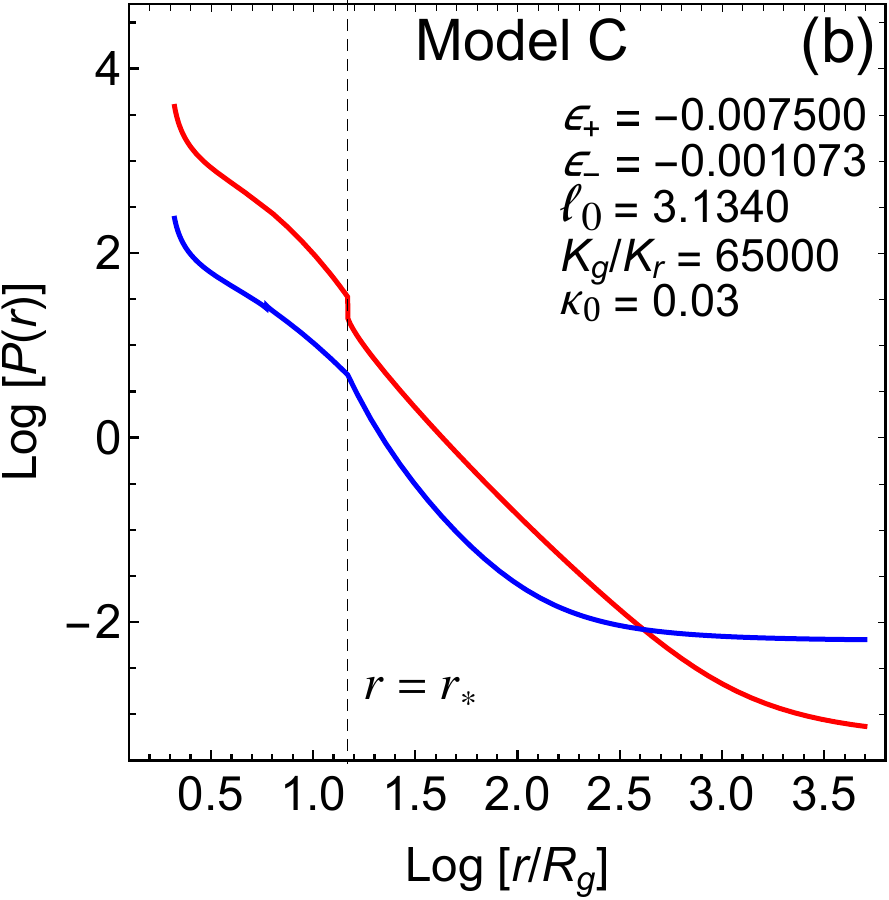}
\centering
\label{fig:1sub2}
\end{subfigure}
\caption{a) Plots of the inflow speed $\vel(r)$ (blue lines) and the effective sound speed $a_{{\rm eff},\kappa}(r)$ (red lines), in units of $c$ for Model~C. The dashed lines denote the one-fluid model of LB05, and the solid lines represent the two-fluid model considered here. b) Plots of the thermal pressure $P_{\rm th}(r)$ (red line) and the relativistic particle pressure $P_{\rm rel}(r)$ (blue line), plotted in cgs units for Model~C.}
\label{figModelCDyn}
\end{figure}

Our computational results show that Model~C yields excellent agreement between the relativistic particle pressure profiles computed using either (i) integration of the relativistic particle distribution, or (ii) solution of the set of hydrodynamical equations (e.g. Figure~\ref{figModelCDyn}b). This establishes the self-consistency of Model C. The dynamical profiles for the two-fluid solution of Model~C are plotted in Figure~\ref{figModelCDyn}a, where the blue and red solid lines represent the bulk flow velocity and the effective sound speed, respectively, and the dashed lines represent the profiles for the corresponding one-fluid model developed by LB05. Note the appearance of the smooth deceleration precursor in the two-fluid velocity profile located just upstream from the shock location, $r_*$, which is qualitatively different from the sharp velocity discontinuity in the one-fluid model (e.g. LB04; LB05). The value obtained for the shock/jet radius in Model~C is $r_*=14.78\,R_g$, which is comparable to the jet-launching radius deduced in the case of M87 by Le et al. (2018).

\begin{figure}
\centering
\begin{subfigure}[b]{.4\textwidth}
\centering
\includegraphics[width=1\linewidth]{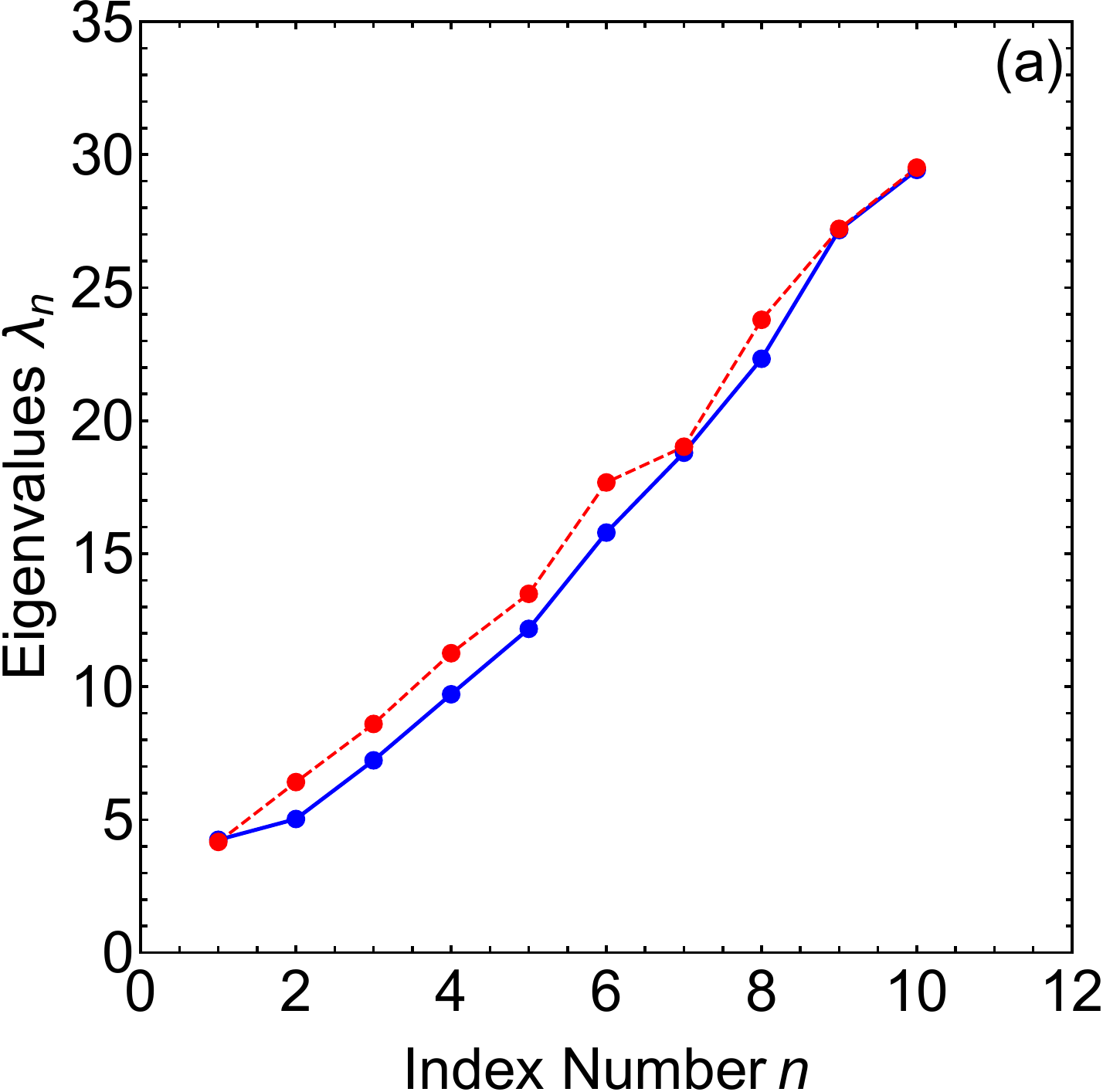}
\label{fig:4sub1}
\end{subfigure}
\quad
\begin{subfigure}[b]{.4\textwidth}
\includegraphics[width=1\linewidth]{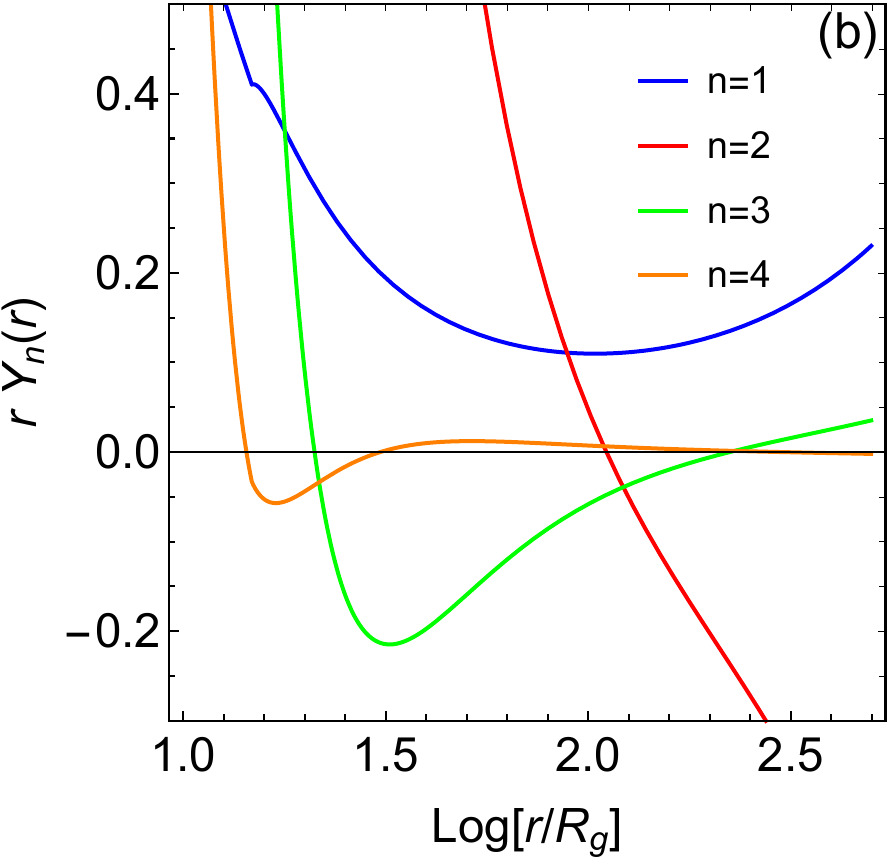}
\centering
\label{fig:4sub2}
\end{subfigure}
\caption{a) Plot of the eigenvalues for our Model~C (blue), compared with the eigenvalues for Model~2 (red) from LB07 (Table \ref{table:table5}). b) Plots of the first four spatial eigenfunctions $Y_n(r)$ for Model~C (equation~\ref{eqYnCases}).}
\label{figMod3EigenvalueFunction}
\end{figure}

The eigenvalues $\lambda_n$ obtained in the application of the two-fluid Model~C to M87 are plotted in Figure~\ref{figMod3EigenvalueFunction}a (blue circles), and compared to those obtained using the one-fluid model in LB07 (red circles). These quantities are also listed in Table \ref{table:table5}. In agreement with LB07, we find that the first eigenvalue, $\lambda_1\sim4$, implying that the particle acceleration process is close to maximum efficiency. This result is consistent with the analogous case of cosmic-ray acceleration (see Blandford \& Ostriker 1978; LB07). Note that the first eigenvalue $\lambda_1$ is also slightly larger in the two-fluid model considered here, as compared to the one-fluid model studied by LB07. This reflects the weakening of the shock acceleration that occurs when the back-reaction of the accelerated particles is included in the dynamical model, as indicated by the deceleration precursor in Figure~\ref{figModelCDyn}a. In Figure~\ref{figMod3EigenvalueFunction}b we plot the solutions for the first four spatial eigenfunctions $Y_n$ (equation~\ref{eqYnCases}), demonstrating that the number of sign changes in $Y_n$ is equal to $n-1$, as expected in the classical Sturm-Liouville problem.

\begin{table}
\centering
\caption{Eigenvalues $\lambda_n$  for our Model~C compared with Model~2 from LB07.}
\begin{tabular}{cccccc}
\hline\hline
$\lambda_n$ & LB07 & Model~C \\
\hline
$\lambda_1$ & 4.165 & 4.244 \\
$\lambda_2$ & 6.415 & 5.027 \\
$\lambda_3$ & 8.600 & 7.232 \\
$\lambda_4$ & 11.259 & 9.715 \\
$\lambda_5$ & 13.491 & 12.175 \\
$\lambda_6$ & 17.678 & 15.793 \\
$\lambda_7$ & 19.022 & 18.796 \\
$\lambda_8$ & 23.792 & 22.327 \\
$\lambda_9$ & 27.211 & 27.166 \\
$\lambda_{10}$ & 29.513 & 29.426 \\
\hline
\end{tabular}
\label{table:table5}
\end{table}

\subsection{Proton Number Conservation Equation}

\begin{figure}
\centering
\begin{subfigure}[b]{.4\textwidth}
\centering
\includegraphics[width=1\linewidth]{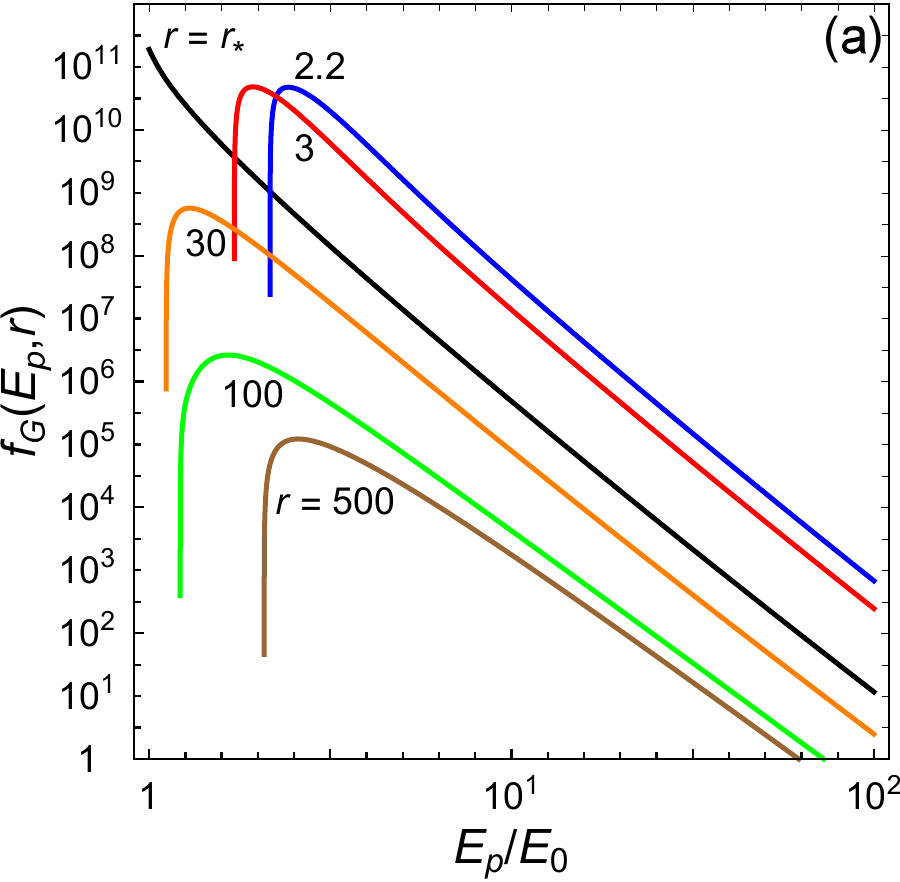}
\label{fig:1sub1b}
\end{subfigure}
\quad
\begin{subfigure}[b]{.41\textwidth}
\includegraphics[width=1\linewidth]{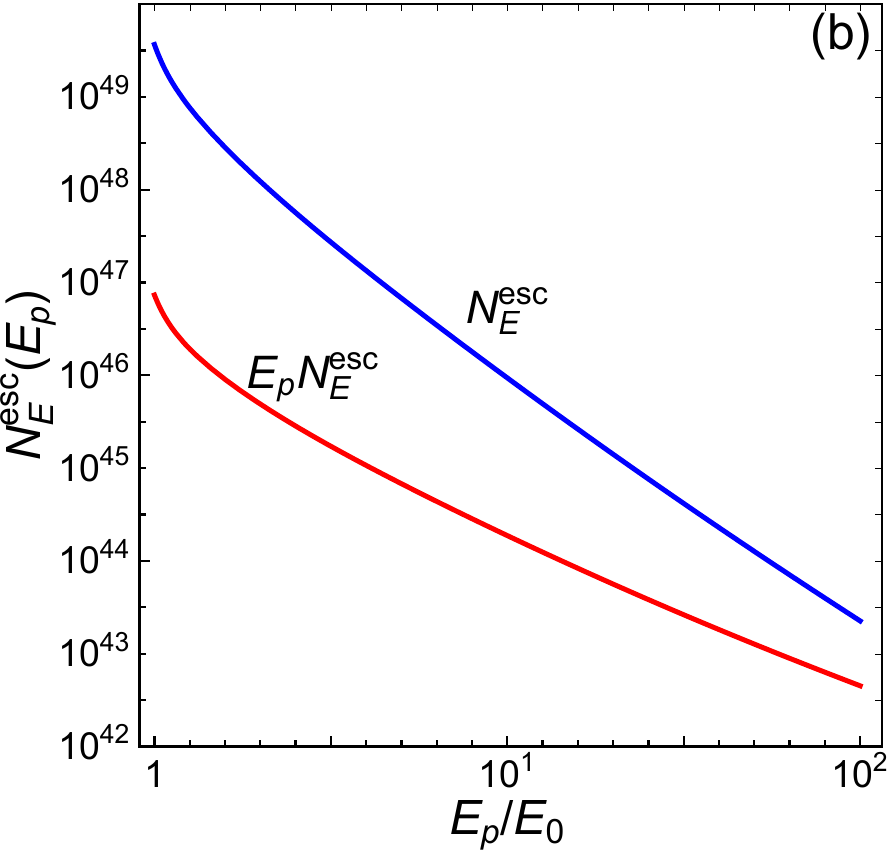}
\centering
\label{fig:1sub2b}
\end{subfigure}
\caption{a) Plots of the solution for the relativistic particle Green's function $\greens(\Eproton,r)$ at various radii inside the disc, in units of ${\rm erg}^{-3}{\rm cm}^{-3}$, computed using equation~(\ref{eqGreensExpansionV1}) for Model~C. b) Plots of the escaping proton energy distribution $\dot N^{\rm esc}_E(\Eproton)$ (blue) and energy distribution $\Eproton\dot N^{\rm esc}_E(\Eproton)$ (red), evaluated using equation~(\ref{eq52LB07}) for Model~C.}
\label{figModelGreensFunction}
\end{figure}

Following the same procedure employed by LB07, applied to the two-fluid model of interest here, we can combine our results for the eigenvalues, eigenfunctions, and expansion coefficients in order to calculate the Green's function, $\greens(\Eproton,r)$, for M87 using the expansion in equation~(\ref{eqGreensExpansionV1}). The resulting Green's function is plotted in Figure~\ref{figModelGreensFunction}a, which depicts $\greens(\Eproton,r)$ as a function of the relativistic proton energy ratio $\Eproton/E_0$ at various radii $r$ in the disc for Model~C from Paper~1, where $E_0=0.002\,$erg is the value of the injected seed proton energy. The corresponding proton injection rate, $\dot N_0$, is computed by ensuring that the power in the injected particles is equal to the power lost from the thermal gas in the disc at the shock location. Note that $\greens=0$ at the injection energy ($\Eproton=E_0$), except at the shock location ($r=r_*$), due to the acceleration of the protons as they diffuse away from the injection radius. In Table \ref{table:table1a} we report the model values obtained for the particle injection rate, $\dot N_0$, the particle escape rate, $\dot N_{\rm esc}$, and the outward and inward particle transport rates in the disc, $\dot N_{\rm I}$ and $\dot N_{\rm II}$, respectively. These various quantities are related via the proton number conservation equation, which requires that
\begin{equation}
\dot N_{\rm I} - \dot N_{\rm II} = \dot N_0 - \dot N_{\rm esc} \ .
\end{equation}
Note that in the inner region ($r < r_*$), the particle transport is in  the inward direction, towards the event horizon, and therefore $\dot N_{\rm II} < 0$. Conversely, in the outer region ($r > r_*$), particles are transported in the outward direction, and therefore $\dot N_{\rm I} > 0$. Furthermore, $|\dot N_{\rm I}| \ll |\dot N_{\rm II}|$, which indicates that the particle distribution is strongly attenuated for $r > r_*$ due to the dominance of inward-bound advection over outward-bound diffusion (see Table \ref{table:table1a}). The attenuation of the particle distribution in the outer region is also apparent in Figure~\ref{figModelGreensFunction}a.

\subsection{Escaping Particle Distribution}
\label{subEscPart}

A primary objective in this study is to characterize the energy distribution of the relativistic protons in the jet escaping from the accretion disc in M87, which generate the observed TeV emission by colliding with ambient protons in clouds or stellar atmospheres in the jet's path. The jet originates as an outflow of plasma blobs containing an isotropic distribution of relativistic protons, surrounded by closed magnetic field lines. Here, we are mainly interested in processes occurring near the base of the jet, since the TeV flare emission seems to be generated when the outflow collides with a cloud or stellar atmosphere located within $\sim 0.01-0.1\,$pc from the black hole. On such small scales, the outflow is only mildly relativistic, with speed $\sim 0.01\,$c (Biretta \& Junor 1995; Junor, Biretta, \& Livio 1999).

We can write down an expression for the energy distribution of the particles escaping from the disc, denoted by $\dot{N}^{\rm esc}_E(\Eproton)$, by integrating equation~(\ref{eqEscape}) with respect to energy and volume, which yields
\begin{equation}
\dot N^{\rm esc}_E(\Eproton) = (4\uppi\, \Eproton)^2r_*H_*cA_0\greens(\Eproton,r_*)
\label{eq52LB07} \ , 
\end{equation}
where $\dot N^{\rm esc}_Ed\Eproton$ denotes the number of protons escaping from the disc per unit time with energy between $\Eproton$ and $\Eproton+d\Eproton$. The protons escape from the disc in the vicinity of the shock, which has a thickness comparable to the magnetic coherence length, $\lambda_{\rm mag} \sim 0.85\,R_g$ in the specific case of Model~C (see Paper~1). The shock thickness is indicated by the extent of the smooth deceleration precursor visible in the plot of the velocity in Figure~\ref{figModelCDyn}. The total number of particles escaping from the disc per second, $\dot N_{\rm esc}$, as well as the total energy escape rate, $L_{\rm esc}$, are computed using the integrals
\begin{equation}
\dot N_{\rm esc} = \int^{\infty}_{E_0}\dot N^{\rm esc}_E(\Eproton)\,d\Eproton = 4\uppi\, r_*H_*cA_0n_* \ , 
\label{eq53LB07}
\end{equation}
and
\begin{equation}
L_{\rm esc} = \int^{\infty}_{E_0}\dot N^{\rm esc}_E(\Eproton) \Eproton\,d\Eproton = 4\uppi\, r_*H_*cA_0U_* \ , 
\label{eq54LB07}
\end{equation}
where $n_*\equiv n_{\rm rel}(r_*)$ and $U_*\equiv U_{\rm rel}(r_*)$ represent the relativistic particle number and energy densities at the shock location, respectively. The values obtained for $L_{\rm esc}$ via equation~(\ref{eq54LB07}) agree well with those listed for $L_{\rm jet}$ in Table 1 from Paper~1, thus confirming that our model satisfies global energy conservation (see equation~\ref{eqDeltaE}).  The escaping particle energy distribution, $\dot N^{\rm esc}_E(\Eproton)$, for Model~C is plotted in Figure~\ref{figModelGreensFunction}b. In Section~\ref{secSecondaryRadiation}, we will use the energy distribution of the escaping protons to compute the secondary radiation spectrum generated when the jet of relativistic protons collides with a cloud or stellar atmosphere in its path.

The computational domain for our problem comprises the base of the jet, where not much collimation or acceleration has yet occurred. On larger scales, the M87 outflow becomes relativistic and is collimated either hydrodynamically or hydromagnetically (see, e.g. Lucchini et al. 2019; Park et al. 2019; Hervet et al. 2017). In order to calculate the asymptotic Lorentz factor of the jet, $\Gamma_{\infty}$, we assign a fluid character to the outflow. As the jet propagation proceeds, the plasma expands and the flow accelerates due to the work done by the plasma blob. This process essentially converts stochastic internal energy of the relativistic particles into directed kinetic energy. We can therefore estimate the asymptotic Lorentz factor of the jet by writing
\begin{equation}
\Gamma_{\infty}=\frac{\langle E_{\rm esc}\rangle}{m_{\rm p} c^2} \ ,
\end{equation}
where the mean energy of the escaping protons is given by
\begin{equation}
\langle E_{\rm esc}\rangle\equiv\frac{U_{\rm rel}(r_*)}{n_{\rm rel}(r_*)} \ .
\end{equation}
The values we obtain for $\Gamma_{\infty}$ are reported in Table \ref{table:table1a}.

\subsection{Self-Consistency of the Number and Energy Density Distributions}
\label{subSelfNEDist}

In Paper~1, we used the two-fluid hydrodynamical model to compute the relativistic proton number and energy density distributions in the accretion disc. This was accomplished by numerically solving second-order ordinary differential equations for the total relativistic particle number density, $n_{\rm rel}(r)$, and the total relativistic particle energy density, $U_{\rm rel}(r)$, given by
\begin{equation}
H \vel_r \frac{dn_{\rm rel}}{dr} = -\frac{n_{\rm rel}}{r}\frac{d}{dr}(rH\vel_r)+\frac{1}{r}\frac{d}{dr}
\left(rH\kappa\frac{dn_{\rm rel}}{dr}\right)+\frac{\dot N_0 \delta(r-r_*)}{4\uppi r_*}
- A_0 c H_* \delta(r-r_*) n_{\rm rel} \ ,
\label{eqVertIntTransNr}
\end{equation}
and 
\begin{equation}
H \vel_r \frac{dU_{\rm rel}}{dr} = -\frac{\gamma_{\rm rel} U_{\rm rel}}{r}\frac{d}{dr}(rH\vel_r)+\frac{1}{r}\frac{d}{dr}\left(rH\kappa\frac{dU_{\rm rel}}{dr}\right) + \frac{\dot N_0 E_0 \delta(r-r_*)}{4\uppi r_*} - A_0 c H_* \delta(r-r_*) U_{\rm rel} \ ,
\label{eqVertIntTransUr}
\end{equation}
respectively.

In the present paper, we have obtained the series solution for the Green's function $\greens(\Eproton,r)$, given by equation~(\ref{eqGreensExpansionV1}). Term-by-term integration of the expansion in equation~(\ref{eqGreensExpansionV1}) yields the expressions (see equations \ref{eqNrUrGreens}) 
\begin{equation}
\begin{split}
n_{\rm rel}(r)\equiv 4\uppi\, E^3_0\sum^{N_{\rm max}}_{n=1}\frac{b_nY_n(r)}{\lambda_n-3} \ , \\
U_{\rm rel}(r)\equiv 4\uppi\, E^4_0\sum^{N_{\rm max}}_{n=1}\frac{b_nY_n(r)}{\lambda_n-4} \ ,
\label{eq51LB07}
\end{split}
\end{equation}
where we generally set $N_{\rm max}=10$ in our applications to M87.

\begin{figure}
\centering
\includegraphics[width=1\linewidth]{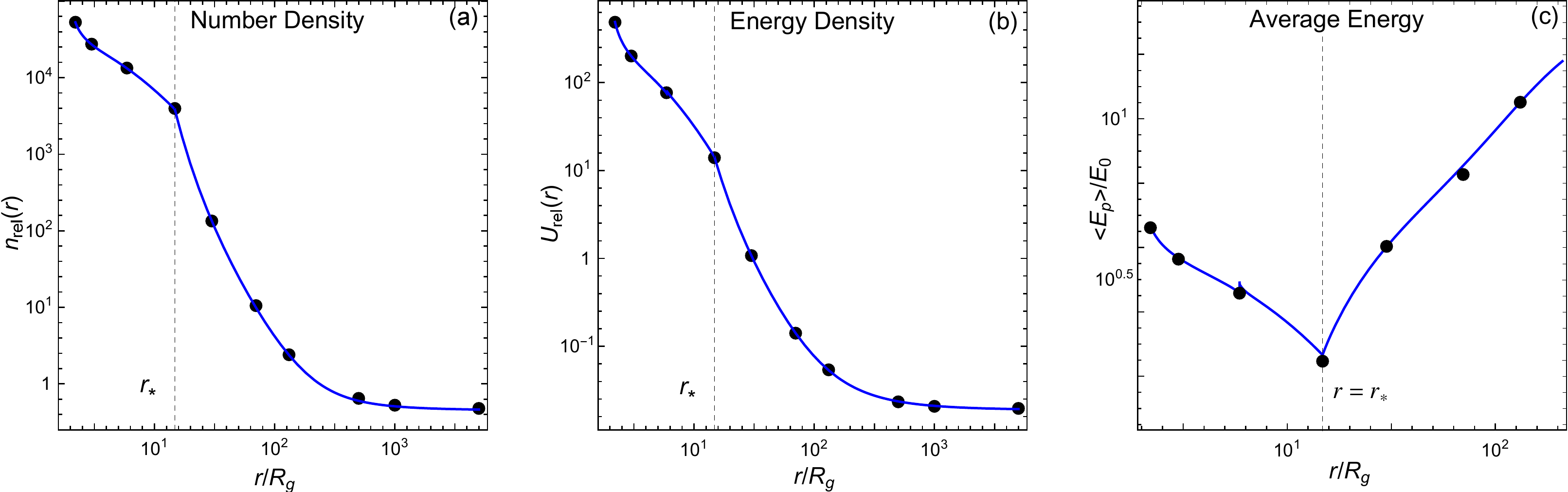}
\caption{Plots of solutions for a) the relativistic particle number density, $n_{\rm rel}(r)$, b) the relativistic particle energy density, $U_{\rm rel}(r)$, and c) the average proton energy, $\langle E_p\rangle/E_0$, computed in cgs units for M87 using Model~C. The solid lines represent the solutions obtained by numerically integrating the differential equations~(\ref{eqVertIntTransNr}) and (\ref{eqVertIntTransUr}), and the filled circles represent the corresponding results obtained via term-by-term integration of the Green's function using equation~(\ref{eq51LB07}). The shock location at $r=r_*$ is indicated. The agreement between the various results confirms the validity of our solution for the Green's function.}
\label{figMeanEnergyDyn}
\end{figure}

With the availability of equations~(\ref{eqVertIntTransNr}), (\ref{eqVertIntTransUr}), and (\ref{eq51LB07}), we have two different ways in which to compute the solutions for the relativistic proton number and energy densities, $n_{\rm rel}(r)$ and $U_{\rm rel}(r)$, respectively. Hence the validity of the solution for the Green's function can be evaluated by comparing the results computed using the two separate methods. The various results for $n_{\rm rel}(r)$ and $U_{\rm rel}(r)$ for Model~C are compared in Figures~\ref{figMeanEnergyDyn}a and \ref{figMeanEnergyDyn}b, respectively, where the solid lines represent the solutions to the second-order equations (\ref{eqVertIntTransNr}) and (\ref{eqVertIntTransUr}), respectively, and the black filled circles represent the results obtained via term-by-term integration of the Green's function (equations~\ref{eq51LB07}). The corresponding profiles for the average proton energy $\langle E_p\rangle(r)$ are plotted in Figure~\ref{figMeanEnergyDyn}c, where
\begin{equation}
\langle E_p\rangle(r) = \frac{U_{\rm rel}(r)}{n_{\rm rel}(r)} \ .
\label{eqErangle}
\end{equation}
The excellent agreement between the profiles of $n_{\rm rel}(r)$, $U_{\rm rel}(r)$, and $\langle E_p\rangle(r)$ computed using equations~(\ref{eqVertIntTransNr}) - (\ref{eqErangle}) confirms the validity of the analysis involved in deriving the Green's function.

\section{Secondary Radiation from Jet-Cloud Interaction}
\label{secSecondaryRadiation}

\begin{figure}
\centering
\includegraphics[width=4in]{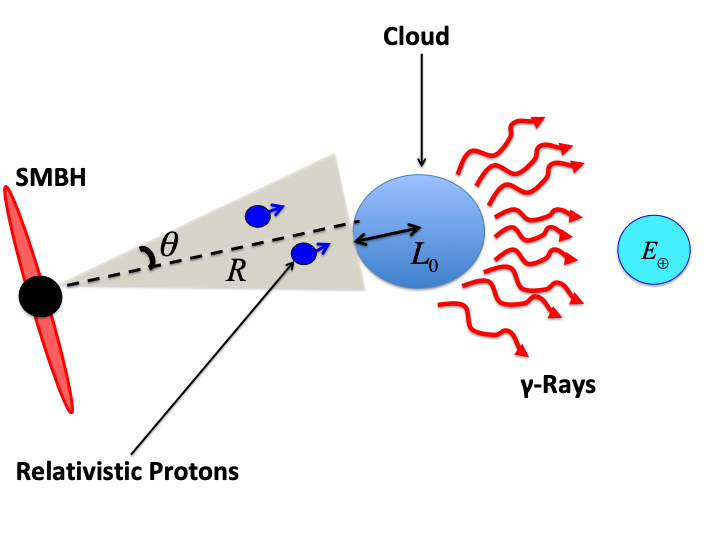}
\caption{Schematic representation of secondary $\gamma$-ray production due to jet of relativistic protons striking a cloud or stellar atmosphere, as seen from Earth.}
\label{figCh5AA}
\end{figure}

In order to compare our theoretical model predictions with the TeV flare data obtained during the high-energy transients observed from M87 in 2004, 2005, and 2010, we need to compute the $\gamma$-ray spectrum produced when the relativistic proton jet collides with the cloud or stellar atmosphere in its path. Collisions between protons in the jet and the cloud produce nuclear reactions that generate secondary radiation via a cascade that begins with muon production and decay. This process has been reviewed by e.g. Eilek \& Kafatos (1983), Barkov et al. (2012), Bj\"ornsson (1999), and Dermer \& Menon (2009). The collision scenario is illustrated in Figure~\ref{figCh5AA}, in which the conical jet has half-angle $\theta$, and the line of sight to Earth is situated within the jet propagation path (e.g. A09). The interaction of the jet with the ambient gas in the cloud generates proton-proton (pp) reactions, resulting in the creation of neutral and charged pions, represented by $\uppi\,^0$, $\uppi\,^+$ and $\uppi\,^-$, respectively. In addition to the TeV emission generated by pp collisions, leptons in the jet will generate SSC emission that contributes to the observed spectrum from radio wavelengths up to GeV energies (Finke et a. 2008).

\subsection{Cross-Field Diffusion from Corona into Outflow}
\label{secLowEnergyMagDef}

As discussed in Section \ref{secIntro1}, the {\it Fermi}-LAT observations of M87 in the GeV energy range obtained in 2008-2009 are not contemporaneous with any of the TeV flares detected by {\it VERITAS} or {\it HESS}. However, following B12 and Fraija \& Marinelli (2016), we will nonetheless use the {\it Fermi}-LAT data to constrain the multi-wavelength fits developed here. In order to avoid over-producing GeV emission beyond the level observed by {\it Fermi}-LAT, B12 introduced a low-energy cutoff in the proton distribution at an energy $E_p\sim1\,$TeV. The low-energy cutoff imposed by B12 was arbitrary, and no associated physical mechanism was suggested. This has motivated us to reconsider the possible physical basis for a low-energy cutoff in the proton distribution striking the cloud or stellar atmosphere. We propose that the low-energy cutoff in the jet proton distribution can be explained as a consequence of energy-dependent cross-field Bohm diffusion from the corona into the blobs of plasma that form the base of the jet outflow. Cross-field diffusion is necessary if the blobs are surrounded by closed magnetic field lines. We provide a quantitative description of this process below.

In the scenario considered here, the protons diffuse vertically out of the disc into the corona according to the prescription worked out in Paper~1, and then subsequently experience cross-field diffusion from the corona into the base of the jet, forming the proton population that eventually collides with the cloud. If the magnetic field lines in the jet outflow are not directly connected with the corona above the accretion disc, then relativistic protons from the corona must enter the base of the jet via cross-field diffusion, which is a process driven by resonant interactions between charged particles and MHD waves (Melrose 1998). Cross-field diffusion occurs when MHD turbulence creates an effective ``wandering'' of the magnetic field lines (Michalek \& Ostrowski 1998). Shalchi \& Dosch (2009) demonstrated that in situations involving strong MHD turbulence, cross-field diffusion approaches the limit of Bohm diffusion, in which the proton mean-free path, $\ell$, is comparable to the particle Larmor radius, $\rL$ (e.g. Kroon et al. 2016), 
\begin{equation}
\rL = \frac{E_p}{q_pB} \ ,
\label{eqLarmor}
\end{equation}
where $q_p$ is the proton charge, $B$ denotes the magnetic field strength, and $E_p$ is the proton energy. In the limit of Bohm diffusion, we can therefore write
\begin{equation}
\ell \sim \rL = 3.34 \times 10^8 \, {\rm cm} \, \left(\frac{E_p}{1\,{\rm TeV}}\right) \left(\frac{B}{10\,{\rm G}}\right)^{-1} \ .
\label{eq3.1}
\end{equation}
The energy threshold for cross-field diffusion depends on the details of the resonance between the protons and the MHD waves propagating in the local magnetic field. We argue below that cross-field diffusion creates a filter that blocks low-energy protons from diffusing into the base of the jet outflow.

The MHD wave distribution is expected to follow a Kraichnan or Kolmogorov wavenumber distribution (Dermer et al. 1996), extending from a maximum driving wavelength, $\lambda_*$, comparable to the disc half-height $H_*$, down through an inertial range, to terminate at a dissipation scale, $\lambda_{\rm diss}$, corresponding to the onset of the particle resonance.
The resonance condition for interactions between Alfv\'en (or magnetosonic) waves and charged particles with velocity $\vel$ and pitch angle cosine $\mu$ can be written as (Miller 1991; Melrose 1998)
\begin{equation}
\omega - k \, \vel \cos\phi \mu \pm n \Omega = 0 \ ,
\label{eq3.4}
\end{equation}
where $\phi$ is the angle between the wave vector and the magnetic field direction, $n$ is the harmonic number, and $\Omega$ denotes the relativistic gyrofrequency, computed using
\begin{equation}
\Omega = \frac{q B}{\gamma_p m_p c} = \frac{c}{\rL} \ .
\label{eq3.5}
\end{equation}
The harmonic number $n = 0$ corresponds to resonance with the parallel electric field, and is associated with magnetosonic waves. On the other hand, the positive integer values $n = 1, 2, 3, \ldots$ correspond to resonances with the transverse electric field, and are associated with Alfv\'en waves.

The dispersion relation for Alfv\'en and magnetosonic waves is given by
\begin{equation}
\omega = \vel_{\rm A} k |\cos\phi| \ ,
\label{eq3.2}
\end{equation}
where $k=2\pi/\lambda$ is the wavenumber and $\vel_{\rm A}$ is the Alfv\'en velocity for a plasma with mass density $\rho$, computed using
\begin{equation}
\vel_{\rm A} = \frac{B}{\sqrt{4 \pi \rho}} \ .
\label{eq3.3}
\end{equation}
For particles with $\vel \gg \vel_{\rm A}$, it follows that $\omega \ll k \vel$, in which case the most important resonance for waves propagating parallel to the field ($\cos\phi=1$) is the cyclotron resonance, with harmonic number $n=1$ (Miller 1991; Michalek \& Ostrowski 1998). In this case, the resonance condition in Equation~(\ref{eq3.4}) reduces to
\begin{equation}
k = \frac{\Omega}{c |\mu|} \ ,
\label{eq3.7a}
\end{equation}
where we have set $\vel=c$ for relativistic particles. Solving for the resonant wavelength yields
\begin{equation}
\lambda = 2 \pi \rL |\mu| \ ,
\label{eq3.9a}
\end{equation}
which can be expressed in terms of the proton energy, $E_p$, as
\begin{equation}
\lambda = \frac{2 \pi |\mu|}{q B} \, E_p
= 2.10 \times 10^9 \, {\rm cm } \, \left(\frac{B}{\rm 10\,G}\right)^{-1} \, \left(\frac{E_p}{1\,\rm TeV}\right) |\mu| \ .
\label{eq3.10}
\end{equation}

A proton with energy $E_p$ and pitch angle cosine $\mu$ will resonate with Alfv\'en waves with wavelength $\lambda$ given by Equation~(\ref{eq3.10}), and will therefore experience cross-field diffusion from the corona into the base of the jet outflow. Setting $\mu = \pm 1$ yields the maximum resonant wavelength for a proton with energy $E_p$, given by
\begin{equation}
\lambda_{\rm max} = \frac{2 \pi}{q B} \, E_p
= 2.10 \times 10^9 \, {\rm cm } \, \left(\frac{B}{\rm 10\,G}\right)^{-1} \, \left(\frac{E_p}{1\,\rm TeV}\right) \ .
\label{eq3.10b}
\end{equation}
The critical proton energy for Bohm diffusion, $E_c$, is therefore obtained by setting $\lambda_{\rm diss} = \lambda_{\rm max}$, where $\lambda_{\rm diss}$ is the dissipation scale for the MHD turbulence. The result obtained for the critical energy is
\begin{equation}
E_c = 0.48 \, {\rm TeV} \, \left(\frac{\lambda_{\rm diss}}{10^9 \, {\rm cm}}\right) \, \left(\frac{B}{\rm 10\,G}\right) \ .
\label{eq3.11}
\end{equation}
Protons with energy $E_p \gtrsim E_c$ will resonate with MHD waves, and will therefore experience cross-field Bohm diffusion from the corona into the base of the jet. Conversely, protons with energy $E_p \lesssim E_c$ will remain in the corona, rather than participating in the jet outflow (Dermer 1988). Since the details of the wave dissipation are not very well understood, we will treat $\lambda_{\rm diss}$ as a free parameter in our numerical calculations. The values $B \sim 10\,$G  and $\lambda_{\rm diss} \sim 10^9\,$cm are comparable to the field strengths and sizes associated with typical solar CME events, and may also be appropriate scales for transients occurring in AGN accretion discs and coronae, if the magnetic field is close to equipartition value (see Paper~1).

Rather than imposing a hard cutoff at the critical proton energy $E_p=E_c$, as employed by B12, we will simulate the stochastic effect of cross-field Bohm diffusion into the base of the outflow by using a smooth low-energy attenuation function given by 
\begin{equation}
F_{\rm Bohm}(E_p) = e^{-E_c/E_p} \ .
\label{eqRoll}
\end{equation}
The utilization of a smooth function, rather than a hard cutoff, is motivated by the fact that Bohm diffusion is fundamentally a three-dimensional random walk, and consequently one does not expect a sharply defined transition energy. The value adopted for the critical energy in our simulations is $E_c=0.624\,$TeV, corresponding to a critical proton Lorentz factor $\gamma_c \sim 665$. The exponential attenuation in the function $F_{\rm Bohm}(E_p)$ at low energies reflects the fact that most of the protons with energy $E_p \lesssim E_c$ are unable to diffuse into the base of the jet, and are confined to the corona.

\subsection{Proton Flux Striking Cloud}
\label{secGammaProduction}

Next we consider the production of $\gamma$-rays resulting from the decay of neutral pions created in collisions between relativistic jet protons and stationary protons in the cloud. The protons in the jet originate in the accretion disc and are energized as a result of Fermi acceleration at the standing shock, located at radius $r_*$. The proton distribution is assumed to be isotropic in the frame of the outflowing plasma blob. In our application, the target cloud (or stellar atmosphere) is located within $\sim 0.01-0.1\,$pc from the central black hole. Biretta \& Junor (1995) have shown that the outflow speed of the M87 jet within this distance range is $\sim 0.01\,$c, and consequently no significant Doppler boosts or relativistic corrections are required in moving between the blob frame and the frame of the target cloud or stellar atmosphere. In this situation, the flux of jet protons, $I_p$, escaping from the disc and striking the cloud at radius $R_c$ from the black hole, is given by
\begin{equation}
I_p(E_p) = \frac{\dot N_E^{\rm esc}(E_p)}{A_{\rm c}}\,F_{\rm Bohm}(E_p)\propto{\rm erg}^{-1}\,{\rm cm}^{-2}\,{\rm s}^{-1} \ , 
\label{eqIp1}
\end{equation}
where $\dot N_E^{\rm esc}$ denotes the energy distribution of the escaping protons (equation~\ref{eq52LB07}), and $F_{\rm Bohm}(E_p)$ is the attenuation function for the proton distribution, due to cross-field Bohm diffusion from the corona into the base of the jet outflow, as described by equation~(\ref{eqRoll}). The factor $A_{\rm c}$ in equation (\ref{eqIp1}) represents the cross-sectional area of the conical jet at radius $R_c$, and is computed using (see Figure~\ref{figCh5AA})
\begin{equation}
A_{\rm c}=\uppi\, R_c^2\theta^2 \ , 
\label{eqACloud}
\end{equation}
where $\theta$ is the half-angle of the jet. Combining equations (\ref{eq52LB07}), (\ref{eqIp1}) and (\ref{eqACloud}) gives for the incident proton flux impinging on the target cloud
\begin{equation}
I_p(E_p) = \frac{16\uppi\,}{R_c^2\theta^2}\,r_* H_* c A_0 E_p^2\greens(E_p,r_*)\,F_{\rm Bohm}(E_p) \ .
\label{eqIpfull}
\end{equation}
The mean energy of the protons in the isotropic distribution striking the cloud, $\langle E_p\rangle_{\rm jet}$, is computed using
\begin{equation}
\langle E_p\rangle_{\rm jet} = \frac{\int_{E_0}^{E_{p,\rm max}} I_p(E_p)E_pdE_p}{\int_{E_0}^{E_{p,\rm max}} I_p(E_p)dE_p} \ ,
\end{equation}
where $I_p(E_p)$ denotes the proton distribution striking the cloud (equation~\ref{eqIpfull}), which includes the filtering effect of cross-field Bohm diffusion through the function $F_{\rm Bohm}(E_p)$, defined in equation~(\ref{eqRoll}). Hence $\langle E_p\rangle_{\rm jet}$ is expected to be much larger than the mean energy of the protons escaping from the disc, denoted by $\langle E_p\rangle$ (equation~\ref{eqErangle}). The mean Lorentz factor of the jet is related to the mean proton energy, $\langle E_p\rangle_{\rm jet}$, via
\begin{equation}
\langle \gamma \rangle_{\rm jet} = \frac{\langle E_p\rangle_{\rm jet}}{m_pc^2} \ .
\label{eqGammaInf}
\end{equation}
The value of $\langle \gamma \rangle_{\rm jet}$ will be computed for each of the numerical examples considered here when we analyze each of the M87 flares observed in 2004, 2005, and 2010.

\begin{figure}
\centering
\includegraphics[width=4in]{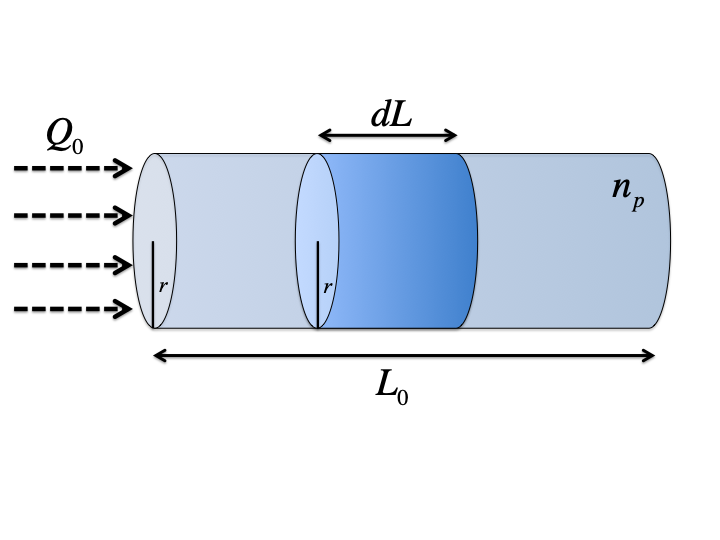}
\caption{Cross-section of a relativistic proton jet with incident flux $Q_0$ propagating through a cloud or stellar atmosphere.}
\label{figCh5B}
\end{figure}

\subsection{Pion Production and Decay}

The TeV $\gamma$-ray energy spectrum produced when the jet of relativistic protons collides with a cloud or stellar atmosphere is generated via inelastic proton-proton (pp) interactions, with subsequent secondary nuclear decay of neutral pions $\uppi\,^0$ and $\eta$ mesons into $\gamma$-rays. The decay scheme for this is generally written as (e.g. Dermer \& Menon 2009)
\begin{equation}
p+p\to\uppi\,^0+X\to2\gamma+X \ , 
\label{eqPPNeutralGamma}
\end{equation}
where $X$ represents all other products generated in the reaction. The secondary interactions can also result in the production of charged pions $\uppi\,^{\pm}$ and other secondaries which decay into high-energy neutrinos, as well as secondary electrons and positrons, all of which can contribute to the SED of an astrophysical source. The charged pions decay according to
\begin{equation}
p+p\to\uppi\,^{\pm}+X\to\mu^{\pm}+X\to e^{\pm}+X \ , 
\label{eqPPNeutralGamma1}
\end{equation}
where $X$ represents additional decay products such as $\nu_{\mu}$, $\bar{\nu}_{\mu}$, $\nu_e$, and $\bar{\nu}_e$. K06 pointed out that $\sim95\%$ of the observed $\gamma$-rays are produced via $\uppi\,^0$ decay rather than $\uppi\,^\pm$ decay, and consequently we shall focus on $\uppi\,^0$ decays here.

The proton-proton interactions are governed by the inelastic cross section (see equation~79 from K06)
\begin{equation}
\sigma_{\uppi}(E_p)=\left(34.3+1.88W+0.25W^2\right)\left[1-\left(\frac{E_{\rm th}}{E_p}\right)^4\right]^2\times10^{-27}\;{\rm cm}^2 \ ,
\label{eqSigInel}
\end{equation}
where $E_p$ is the energy of the proton, $E_{\rm th}=1.22\times10^{-3}$ TeV is the threshold energy for production of ${\uppi}^0$ mesons, and
\begin{equation}
W=\ln\left(\frac{E_p}{1\,{\rm TeV}}\right) \ .
\label{eqLnL}
\end{equation}
Equation~(\ref{eqSigInel}) represents an approximate numerical fit to the data in the SIBYLL code, and is valid for both high and low proton energies (K06).

In order to compute the $\gamma$-ray spectrum resulting from pion decay, we must first describe the geometry of the jet-cloud interaction. Figure~\ref{figCh5B} depicts the jet colliding with a cloud of radius $L_0$, where $Q_0$ represents the incident flux of jet protons and $n_p$ denotes the target proton number density. The probability that an incident proton with energy $\Eproton$ in the jet will collide with a target proton in the cloud to produce a pion in the differential distance between $L$ and $L+dL$ is equal to
\begin{equation}
d{\cal P}_{\uppi} = n_p\sigma_{\uppi}(E_p)\,dL \ ,
\label{eqdP}
\end{equation}
where $\sigma_{\uppi}(E_p)$ is the cross section for pp pion production (equation~\ref{eqSigInel}). It follows that the flux of protons, $Q$, penetrating the cloud is exponentially attenuated by pion production, so that we have
\begin{equation}
Q(E_p, L) = Q_0\,e^{-n_p\sigma_{\uppi}(E_p)L} \ ,
\end{equation}
where $Q_0$ is the incident proton flux impinging on the cloud. The survival probability for protons to cross to the other side of the spherical cloud without producing a pion can therefore be estimated using
\begin{equation}
{\cal P}_{\rm survive}(E_p) = \frac{Q(\Eproton,L_0)}{Q_0}=e^{-n_p\sigma_{\uppi}(E_p)L_0} \ ,
\end{equation}
and consequently, the probability that an incident proton will produce a pion somewhere inside the cloud is equal to 
\begin{equation}
{\cal P}_{\uppi}(E_p) = 1-{\cal P}_{\rm survive}=1-e^{-n_p\sigma_{\uppi}(E_p)L_0} \ .
\label{eqPsurv1}
\end{equation}
For typical cloud densities and sizes, at high proton energies, $n_p\sigma_{\uppi}(E_p)L_0\ll1$, and therefore we can linearize equation (\ref{eqPsurv1}) to obtain for the pion production probability per incident proton
\begin{equation}
{\cal P}_{\uppi}(E_p)=n_p\sigma_{\uppi}(E_p) L_0 \ .
\label{eqPsurv2}
\end{equation}

\subsection{Gamma-ray Spectrum}

Next we calculate the $\gamma$-ray spectrum observed at Earth as a result of neutral pion decays generated when the jet of relativistic protons collides with a cloud or stellar atmosphere. Formally, the specific $\gamma$-ray energy flux observed at Earth is given by
\begin{equation}
F_E(E) = \left(\frac{R_c}{D}\right)^2 F'_E (E) \propto{\rm erg}\;{\rm TeV}^{-1}\;{\rm cm}^{-2}\;{\rm s}^{-1} \ ,
\label{eqEarthGamma1}
\end{equation}
where $D$ is the distance from the black hole to Earth, $R_c$ is the distance from the black hole to the target cloud, and $F'_E$ is the specific $\gamma$-ray flux measured in the frame of the cloud. The $\gamma$-ray flux measured in the frame of the cloud due to contributions from all incident proton energies, $\Eproton$, is computed using
\begin{equation}
F'_E(E) = \int \frac{dF'_E}{d\Eproton}\,d\Eproton \ ,
\label{eqdFedE1a}
\end{equation}
where $dF'_E/d\Eproton$ is the differential $\gamma$-ray flux due to protons with a specific energy $\Eproton$. As discussed in detail by K06, there are two energy regimes of interest, depending on whether the energy $E$ of the outgoing $\gamma$-ray produced at the end of the cascade is above or below 0.1\,TeV. We treat each of these cases separately below.

\subsubsection{Gamma-ray Energy $E\geq0.1\,$TeV}
In the frame of the cloud, the differential $\gamma$-ray flux, due to protons with a specific energy $\Eproton$, is given by (K06)
\begin{equation}
\frac{dF'_E}{dE_p} = I_p(E_p){\cal P}_{\uppi}(E_p)\frac{E}{E_p}F_{\rm K}\left(\frac{E}{E_p},E_p\right) \propto {\rm TeV}^{-1}\;{\rm cm}^{-2}\;{\rm s}^{-1} \ .
\label{eqdFedE1}
\end{equation}
Here, $F_{\rm K}$ represents the number of photons produced in the dimensionless photon energy interval $(x,x+dx)$ per proton collision, where $x=E/\Eproton$. Based on fits to results obtained using the SIBYLL code (see equation~58 of K06), $F_{\rm K}$ can be approximated in the $\gamma$-ray energy range $0.1\,{\rm TeV}\leq E_p\leq10^5\,{\rm TeV}$ using the analytical expression
\begin{equation}
F_{\rm K}(x,E_p)=B\frac{\ln(x)}{x}\left[\frac{1-x^{\beta}}{1+kx^{\beta}(1-x^{\beta})}\right]^4\left[\frac{1}{\ln(x)}-\frac{4\beta x^{\beta}}{1-x^{\beta}}-\frac{4k\beta x^{\beta}(1-2x^{\beta})}{1+kx^{\beta}(1-x^{\beta})}\right] \ ,
\label{eqFK}
\end{equation}
where the parameters $B,\,\beta$ and $k$ are defined by
\begin{equation}
B=1.30+0.14W+0.011W^2 \ ,
\end{equation}
\begin{equation}
\beta=\frac{1}{1.7+0.11W+0.008W^2} \ ,
\end{equation}
\begin{equation}
k=\frac{1}{0.801+0.049W+0.014W^2} \ ,
\end{equation}
and $W$ is computed using equation~(\ref{eqLnL}). By combining equations (\ref{eqdFedE1a}) and (\ref{eqdFedE1}), we find that the $\gamma$-ray flux in the frame of the cloud for photon energies $E\geq 0.1\,{\rm TeV}$ is given by
\begin{equation}
F'_E(E) = \int_{E}^{E_{p,{\rm max}}}I_p(E_p){\cal P}_{\uppi}(E_p)\frac{E}{E_p}F_{\rm K}\left(\frac{E}{E_p},E_p\right)\,d\Eproton\ ,\qquad E\geq0.1\,{\rm TeV} \ ,
\label{eqPhiHigh}
\end{equation}
where the upper bound $E_{p,\rm max}$ is the maximum proton energy, which is treated as a free parameter, as discussed in Section \ref{secAstroApps}.

\subsubsection{Gamma-ray Energy $E\leq0.1\,$TeV}
Following the development in K06, we find that a different approach needs to be utilized in order to compute the observed $\gamma$-ray spectrum for the case with photon energy $E<0.1\,{\rm TeV}$. In this case, the pion distribution function has a delta-function dependence on the proton kinetic energy, $E_{\rm kin}=E_p-m_pc^2$, and therefore we can write the pion energy as (see equation~75 from K06)
\begin{equation}
E_{\uppi}=K_{\uppi}E_{\rm kin}=K_{\uppi}(E_p-m_pc^2) \ ,
\label{eqEpiEp}
\end{equation}
where the constant $K_{\pi}=0.17$, based on fits to the output from the SYBILL code at low energies, $1\,{\rm GeV}\leq E\leq0.1\,{\rm TeV}$. In the low-energy regime, we therefore find that Equation (\ref{eqdFedE1}) can be written as
\begin{equation}
\frac{dF'_E}{dE_p} = \frac{2\,E}{\sqrt{E_{\uppi\,}^2-m^2_{\uppi}c^4}}\,I_p(E_p){\cal P}_{\uppi}(E_p) \propto {\rm TeV}^{-1}\;{\rm cm}^{-2}\;{\rm s}^{-1} \ ,
\label{eqdFedE1b}
\end{equation}
where the factor $2/(E_{\uppi\,}^2-m^2_{\uppi}c^4)^{1/2}$ represents the $\gamma$-ray distribution resulting from neutral pion decay (see equation 1-88 from Stecker 1971). 

By analogy with equation (\ref{eqPhiHigh}), we conclude that the flux of $\gamma$-rays with energy $E\leq0.1\,$TeV measured in the frame of the cloud, resulting from collisions with jet protons of all energies, is given by the integral
\begin{equation}
F'_E(E) = \int_{E_{p,{\rm min}}}^{E_{p,{\rm max}}}\frac{2\,E}{\sqrt{E_{\uppi\,}^2-m^2_{\uppi}c^4}}\,I_p(E_p){\cal P}_{\uppi}(E_p)\,d\Eproton\ ,\qquad E\leq0.1\,{\rm TeV} \ ,
\label{eqPhiLow1}
\end{equation}
where $E_{\uppi}$ is evaluated as a function of $\Eproton$ using equation (\ref{eqEpiEp}), and the upper energy bound, $E_{p,\rm max}$ is a free parameter. The low-energy bound, $E_{p,\rm min}$, in equation (\ref{eqPhiLow1}) is defined by
\begin{equation}
E_{p,\rm min}=\frac{E_{\uppi,\rm min}}{K_{\uppi\,}}+m_pc^2 \ ,
\label{eqEpMin}
\end{equation}
where $E_{\uppi, \rm min}=E+m^2_{\uppi}c^4/(4E)$ is the minimum pion energy required to produce a $\gamma$-ray with energy $E$ (Stecker 1971).

We are now in a position to compute the $\gamma$-ray spectrum observed at Earth, denoted by $F_E(E)$. By combining equations (\ref{eqIpfull}), (\ref{eqPsurv2}), (\ref{eqEarthGamma1}), (\ref{eqdFedE1a}), (\ref{eqPhiHigh}), and (\ref{eqPhiLow1}), we find that in the photon energy range $E\geq0.1\,{\rm TeV}$, the observed $\gamma$-ray spectrum is given by
\begin{equation}
F_E(E) = \frac{16\uppi\,\xi}{D^2}\,r_* H_* c A_0\,E \int_{E}^{E_{p,{\rm max}}} E_p\greens(E_p,r_*)\,F_{\rm Bohm}(E_p)\sigma_{\uppi}(E_p)F_{\rm K}\left(\frac{E}{E_p},E_p\right)\,d\Eproton\ ,\qquad E\geq0.1\,{\rm TeV} \ ,
\label{eqPhiHighv2}
\end{equation}
and in the photon energy range $E\leq0.1\,{\rm TeV}$, the spectrum $F_E(E)$ is given by
\begin{equation}
F_E(E) = \frac{32\uppi\,\xi}{D^2}\,\,r_* H_* c A_0\,E \int_{E_{p,{\rm min}}}^{E_{p,{\rm max}}} \frac{E_p^2}{\sqrt{E_{\uppi\,}^2-m^2_{\uppi}c^4}}\greens(E_p,r_*)\,F_{\rm Bohm}(E_p)\sigma_{\uppi}(E_p)\,d\Eproton\ ,\qquad E\leq0.1\,{\rm TeV} \ ,
\label{eqPhiLow1v2}
\end{equation}
where $E_{\uppi}$ is computed using equation (\ref{eqEpiEp}), the lower bound $E_{p,\rm min}$ is computed using equation (\ref{eqEpMin}), and the upper bound $E_{p,\rm max}$ is a free parameter. The similarity parameter $\xi$ appearing in equations~(\ref{eqPhiHighv2}) and (\ref{eqPhiLow1v2}) is defined by
\begin{equation}
\xi \equiv \frac{\Psi}{\theta^2} \propto {\rm cm}^{-2} \ ,
\label{eqXiMod}
\end{equation}
where $\theta$ denotes the half-angle of the proton jet, and the column density of the cloud or stellar atmosphere, $\Psi$, is computed using
\begin{equation}
\Psi \equiv n_p L_0 \ .
\label{eqPsiMod}
\end{equation}
It is interesting to note that the observed $\gamma$-ray flux, $F_E(\Eproton)$, depends on the cloud density $n_p$, the cloud radius $L_0$, and the half-angle $\theta$ only through the similarity parameter $\xi$, which is varied in order to obtain acceptable fits to the observed TeV $\gamma$-ray spectra.

\section{APPLICATION TO M87}
\label{secAstroApps}
\begin{table}
\centering
\caption{Model parameters for comparison with the 2010 {\it VERITAS} data, with $\xi=6.21\times10^{25}\,{\rm cm}^{-2}$.}
\begin{tabular}[width=1\columnwidth]{ccccccc}
\hline\hline
$\theta\,(^{\circ})$ & $\Psi\,({\rm cm}^{-2})$ & $L_0\,({\rm cm})$ & $n_p\,({\rm cm}^{-3})$ & $\Delta t$ (days) & $r_j\,({\rm cm})$ & $z_c\,({\rm cm})$\\
\hline
2 & $7.56\times10^{22}$ & $10^{13}$ & $7.56\times10^9$ & 5 & $1.12\times10^{15}$ & $3.21\times10^{16}$\\
$\hdots$ &$\hdots$ & $\hdots$ & $\hdots$ & 6 & $1.27\times10^{15}$ &  $3.62\times10^{16}$\\
$\hdots$ &$\hdots$ & $\hdots$ &$\hdots$ & 7 &$1.40\times10^{15}$ &  $4.01\times10^{16}$\\
$\hdots$ &$\hdots$ & $10^{14}$ & $7.56\times10^8$ & 5 & $1.12\times10^{15}$ & $3.21\times10^{16}$\\
$\hdots$ &$\hdots$ & $\hdots$ & $\hdots$ & 6 & $1.27\times10^{15}$ &  $3.62\times10^{16}$\\
$\hdots$ &$\hdots$ & $\hdots$ & $\hdots$ & 7 & $1.40\times10^{15}$ &  $4.01\times10^{16}$\\
\phantom{ }\\
10 & $1.89\times10^{24}$ & $10^{13}$ & $1.89\times10^{11}$ & 5 & $1.92\times10^{15}$ & $1.08\times10^{16}$\\
$\hdots$ &$\hdots$ & $\hdots$ & $\hdots$ & 6 & $2.16\times10^{15}$ & $1.22\times10^{16}$\\
$\hdots$ &$\hdots$ & $\hdots$ & $\hdots$ & 7 & $2.40\times10^{15}$ & $1.35\times10^{16}$\\
$\hdots$ & $\hdots$ & $10^{14}$ & $1.89\times10^{10}$ & 5 & $1.92\times10^{15}$ & $1.08\times10^{16}$\\
$\hdots$ & $\hdots$ & $\hdots$ & $\hdots$ & 6 & $2.16\times10^{15}$ & $1.22\times10^{16}$\\
$\hdots$ & $\hdots$ & $\hdots$ & $\hdots$ & 7 & $2.40\times10^{15}$ & $1.35\times10^{16}$\\
\phantom{ }\\
17 & $5.46\times10^{24}$ & $10^{13}$ & $5.46\times10^{11}$ & 5 & $2.29\times10^{15}$ & $7.37\times10^{15}$\\
$\hdots$ & $\hdots$  & $\hdots$ & $\hdots$ & 6 & $2.58\times10^{15}$ & $8.32\times10^{15}$\\
$\hdots$ & $\hdots$  & $\hdots$ & $\hdots$ & 7 & $2.86\times10^{15}$ & $9.22\times10^{15}$\\
$\hdots$ & $\hdots$  & $10^{14}$ & $5.46\times10^{10}$ & 5 & $2.29\times10^{15}$ & $7.37\times10^{15}$\\
$\hdots$ & $\hdots$  & $\hdots$ & $\hdots$ & 6 & $2.58\times10^{15}$ & $8.32\times10^{15}$\\
$\hdots$ & $\hdots$  & $\hdots$ & $\hdots$ & 7 & $2.86\times10^{15}$ & $9.22\times10^{15}$\\
\hline
\end{tabular}
\label{table:tableNew1} 
\end{table}
\begin{table}
\centering
\caption{Model parameters for comparison with the 2005 {\it HESS} data, with $\xi=1.56\times10^{25}\,{\rm cm}^{-2}$.}
\begin{tabular}[width=1\columnwidth]{ccccccc}
\hline\hline
$\theta\,(^{\circ})$ & $\Psi\,({\rm cm}^{-2})$ & $L_0\,({\rm cm})$ & $n_p\,({\rm cm}^{-3})$ & $\Delta t$ (days) & $r_j\,({\rm cm})$ & $z_c\,({\rm cm})$\\
\hline
2 & $1.90\times10^{22}$ & $10^{13}$ & $1.90\times10^9$ & 5 & $1.12\times10^{15}$ & $3.21\times10^{16}$\\
$\hdots$ &$\hdots$ & $\hdots$ & $\hdots$ & 6 & $1.27\times10^{15}$ &  $3.62\times10^{16}$\\
$\hdots$ &$\hdots$ & $\hdots$ &$\hdots$ & 7 &$1.40\times10^{15}$ &  $4.01\times10^{16}$\\
$\hdots$ &$\hdots$ & $10^{14}$ & $1.90\times10^8$ & 5 & $1.12\times10^{15}$ & $3.21\times10^{16}$\\
$\hdots$ &$\hdots$ & $\hdots$ & $\hdots$ & 6 & $1.27\times10^{15}$ &  $3.62\times10^{16}$\\
$\hdots$ &$\hdots$ & $\hdots$ & $\hdots$ & 7 & $1.40\times10^{15}$ &  $4.01\times10^{16}$\\
\phantom{ }\\
10 & $4.74\times10^{23}$ & $10^{13}$ & $4.74\times10^{10}$ & 5 & $1.92\times10^{15}$ & $1.08\times10^{16}$\\
$\hdots$ &$\hdots$ & $\hdots$ & $\hdots$ & 6 & $2.16\times10^{15}$ & $1.22\times10^{16}$\\
$\hdots$ &$\hdots$ & $\hdots$ & $\hdots$ & 7 & $2.40\times10^{15}$ & $1.35\times10^{16}$\\
$\hdots$ & $\hdots$ & $10^{14}$ & $4.74\times10^{9}$ & 5 & $1.92\times10^{15}$ & $1.08\times10^{16}$\\
$\hdots$ & $\hdots$ & $\hdots$ & $\hdots$ & 6 & $2.16\times10^{15}$ & $1.22\times10^{16}$\\
$\hdots$ & $\hdots$ & $\hdots$ & $\hdots$ & 7 & $2.40\times10^{15}$ & $1.35\times10^{16}$\\
\phantom{ }\\
17 & $1.37\times10^{24}$ & $10^{13}$ & $1.37\times10^{11}$ & 5 & $2.29\times10^{15}$ & $7.37\times10^{15}$\\
$\hdots$ & $\hdots$  & $\hdots$ & $\hdots$ & 6 & $2.58\times10^{15}$ & $8.32\times10^{15}$\\
$\hdots$ & $\hdots$  & $\hdots$ & $\hdots$ & 7 & $2.86\times10^{15}$ & $9.22\times10^{15}$\\
$\hdots$ & $\hdots$  & $10^{14}$ & $1.37\times10^{10}$ & 5 & $2.29\times10^{15}$ & $7.37\times10^{15}$\\
$\hdots$ & $\hdots$  & $\hdots$ & $\hdots$ & 6 & $2.58\times10^{15}$ & $8.32\times10^{15}$\\
$\hdots$ & $\hdots$  & $\hdots$ & $\hdots$ & 7 & $2.86\times10^{15}$ & $9.22\times10^{15}$\\
\hline
\end{tabular}
\label{table:tableNew2} 
\end{table}
\begin{table}
\centering
\caption{Model parameters for comparison with the 2004 {\it HESS} data, with $\xi=3.11\times10^{24}\,{\rm cm}^{-2}$.}
\begin{tabular}[width=1\columnwidth]{ccccccc}
\hline\hline
$\theta\,(^{\circ})$ & $\Psi\,({\rm cm}^{-2})$ & $L_0\,({\rm cm})$ & $n_p\,({\rm cm}^{-3})$ & $\Delta t$ (days) & $r_j\,({\rm cm})$ & $z_c\,({\rm cm})$\\
\hline
2 & $3.78\times10^{21}$ & $10^{13}$ & $3.78\times10^8$ & 5 & $1.12\times10^{15}$ & $3.21\times10^{16}$\\
$\hdots$ &$\hdots$ & $\hdots$ & $\hdots$ & 6 & $1.27\times10^{15}$ &  $3.62\times10^{16}$\\
$\hdots$ &$\hdots$ & $\hdots$ &$\hdots$ & 7 &$1.40\times10^{15}$ &  $4.01\times10^{16}$\\
$\hdots$ &$\hdots$ & $10^{14}$ & $3.78\times10^7$ & 5 & $1.12\times10^{15}$ & $3.21\times10^{16}$\\
$\hdots$ &$\hdots$ & $\hdots$ & $\hdots$ & 6 & $1.27\times10^{15}$ &  $3.62\times10^{16}$\\
$\hdots$ &$\hdots$ & $\hdots$ & $\hdots$ & 7 & $1.40\times10^{15}$ &  $4.01\times10^{16}$\\
\phantom{ }\\
10 & $9.46\times10^{22}$ & $10^{13}$ & $9.46\times10^{9}$ & 5 & $1.92\times10^{15}$ & $1.08\times10^{16}$\\
$\hdots$ &$\hdots$ & $\hdots$ & $\hdots$ & 6 & $2.16\times10^{15}$ & $1.22\times10^{16}$\\
$\hdots$ &$\hdots$ & $\hdots$ & $\hdots$ & 7 & $2.40\times10^{15}$ & $1.35\times10^{16}$\\
$\hdots$ & $\hdots$ & $10^{14}$ & $9.46\times10^{8}$ & 5 & $1.92\times10^{15}$ & $1.08\times10^{16}$\\
$\hdots$ & $\hdots$ & $\hdots$ & $\hdots$ & 6 & $2.16\times10^{15}$ & $1.22\times10^{16}$\\
$\hdots$ & $\hdots$ & $\hdots$ & $\hdots$ & 7 & $2.40\times10^{15}$ & $1.35\times10^{16}$\\
\phantom{ }\\
17 & $2.73\times10^{23}$ & $10^{13}$ & $2.73\times10^{10}$ & 5 & $2.29\times10^{15}$ & $7.37\times10^{15}$\\
$\hdots$ & $\hdots$  & $\hdots$ & $\hdots$ & 6 & $2.58\times10^{15}$ & $8.32\times10^{15}$\\
$\hdots$ & $\hdots$  & $\hdots$ & $\hdots$ & 7 & $2.86\times10^{15}$ & $9.22\times10^{15}$\\
$\hdots$ & $\hdots$  & $10^{14}$ & $2.73\times10^{9}$ & 5 & $2.29\times10^{15}$ & $7.37\times10^{15}$\\
$\hdots$ & $\hdots$  & $\hdots$ & $\hdots$ & 6 & $2.58\times10^{15}$ & $8.32\times10^{15}$\\
$\hdots$ & $\hdots$  & $\hdots$ & $\hdots$ & 7 & $2.86\times10^{15}$ & $9.22\times10^{15}$\\
\hline
\end{tabular}
\label{table:tableNew3} 
\end{table}
\begin{figure}
\centering
\includegraphics[width=1\linewidth]{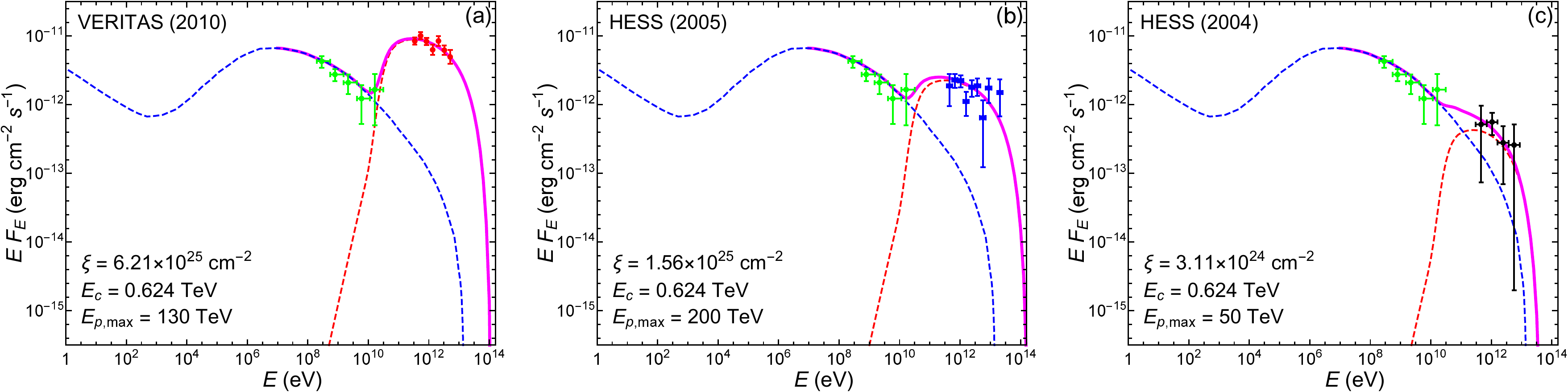}
\caption{Plots of our theoretical TeV radiation spectrum (equations~\ref{eqPhiHighv2} and \ref{eqPhiLow1v2}, red dashed lines), combined with the leptonic SSC model from A09 (blue dashed lines). The total theoretical spectra are indicated by the solid magenta lines. The theoretical spectra are compared with the {\it Fermi}-LAT data from A09 and the TeV spectra detected by a) {\it VERITAS} (2010), b) {\it HESS} (2005), and c) {\it HESS} (2004).}
\label{figCh5C}
\end{figure}

The theoretical framework developed in the preceding sections can now be used to compute the spectrum of secondary TeV $\gamma$-rays generated when the jet of relativistic protons collides with a cloud or stellar atmosphere, and the results can be compared with the observations of M87 obtained using {\it HESS} and {\it VERITAS}. The calculations we perform here are based on Model~C from Paper~1, for which we set the black hole mass $M=6.5\times 10^9\,\msun$ (Akiyama et al. 2019), the accretion rate $\dot{M}=1.51\times10^{-1}\,\msun\,\text{yr}^{-1}$, the jet luminosity $L_{\rm jet}=5.5\times10^{43}\,\text{erg s}^{-1}$, the shock radius $r_*=14.780$, the entropy ratio $K_{\rm th}/K_{\rm rel}=65,000$, the diffusion coefficient $\kappa_0=0.03$, the upstream energy transport rate $\epsilon_-=-0.001073$, and the specific angular momentum $\ell_0=3.1340$ (see Tables \ref{table:table1} and \ref{table:table1a}).

With the dynamical model for the disc and the jet outflow determined as described above, there are few additional free parameters that also need to be specified in order to compute the TeV $\gamma$-ray spectrum produced when the jet collides with the cloud or stellar atmosphere. These quantities are the similarity parameter, $\xi$, the critical proton energy for cross-field Bohm diffusion, $E_c$, and the maximum proton energy, $E_{p,\rm max}$. The values of these parameters are varied in order to obtain reasonable fits to the $\gamma$-ray spectra detected by {\it HESS} and {\it VERITAS} during the flares observed in 2004, 2005, and 2010. In addition to the requirement of matching the observed $\gamma$-ray spectra, there are also additional self-consistency constraints on the parameters, related to the observed variability timescale, and the necessity of locating the cloud or stellar atmosphere above the disc half-thickness $H_*$. We explore the implications of these additional constraints below.

Following the interpretation of B12, we posit that the variability timescale, $\Delta t \sim\,$several days, associated with the $\gamma$-ray transients observed in 2004, 2005 and 2010 is due to the passage of a cloud or stellar atmosphere, moving through the jet funnel with the local Keplerian velocity at a radius $R_c$ from the central black hole. We note that the variability in the B12 scenario is partly due to the evolution of the cloud, due to the absorption of energy from the jet. In our approach, the evolution of the cloud properties is neglected, but we expect that this process would not have a significant effect on the variability properties of the transient $\gamma$-ray emission considered here. The variability timescale is therefore computed using
\begin{equation}
\Delta t = \frac{2\,r_j}{V_{\rm Kep}} \ ,
\label{eqVariabilityTime}
\end{equation}
where the radius of the jet funnel, $r_j$, is given by
\begin{equation}
r_j = \theta R_c \ ,
\label{eqRc1}
\end{equation}
and the local Keplerian velocity is defined by
\begin{equation}
V_{\rm Kep} = \left(\frac{GM}{R_c}\right)^{1/2} \ .
\label{eqRc2}
\end{equation}
By combining equations~(\ref{eqVariabilityTime}), (\ref{eqRc1}) and (\ref{eqRc2}), we obtain
\begin{equation}
\Delta t = \frac{2\,\theta}{\sqrt{GM}}\,R^{3/2}_c \ .
\label{eqRc3}
\end{equation}
Hence, the requirement of a variability timescale $\Delta t\sim$ a few days creates a constraint on the two parameters $R_c$ and $\theta$ which must be satisfied in our attempts to fit the observed high-energy spectra.

We must also ensure that the cloud or stellar atmosphere is located outside the vertically extended accretion disc. This geometrical condition can be written as
\begin{equation}
z_c>H_* \ ,
\label{eqZcHCondition}
\end{equation}
where the altitude of the cloud above the disc plane, $z_c$, is given by
\begin{equation}
z_c = R_c\cos\theta \ ,
\label{eqZcRc}
\end{equation}
and $H_*$ is the half-thickness of the disc at the shock radius, $r_*$, which is the point of origin of the jet outflow. We note that based on the value of $H_*$ stated in gravitational units in Table \ref{table:table1}, combined with the black hole mass for M87 $(M=6.5\times10^9\,M_{\odot})$, we obtain in cgs units $H_*=7.19\times10^{15}\,{\rm cm}$. In addition, we must also require that the radius of the jet funnel, $r_j$, is larger than the cloud radius, $L_0$, 
\begin{equation}
r_j>L_0 \ ,
\label{eqrjL0Condition}
\end{equation}
in order to ensure that the variability timescale is correctly computed using equation (\ref{eqVariabilityTime}). We have confirmed that all of these constraints are satisfied for all of the calculations performed here, as documented in Tables~\ref{table:tableNew1}, \ref{table:tableNew2}, and \ref{table:tableNew3} for the 2010 {\it VERITAS}, 2005 {\it HESS}, and 2004 {\it HESS} data sets, respectively.

\subsection{Constructing the SED}

The hadronic model developed here is capable of reproducing the {\it VERITAS} and {\it HESS} data for the 2004, 2005, and 2010 TeV flares observed from M87. However, in order to develop an integrated physical description of the source, we need to compute the complete multi-wavelength spectrum, extending from radio wavelengths up the TeV energies. We can accomplish this by creating a superposition that combines our model with the one-zone leptonic SSC jet model of Finke et al. (2008). We argue that a superposition of the two models is reasonable since the radiating particle populations in the two scenarios are separate. Following A09 and Fraija \& Marinelli (2016), we will use the {\it Fermi}-{\rm LAT} data to constrain the multi-wavelength fits, and we will demonstrate that a superposition of our model with the SSC model of Finke et al. (2008) is able to fit the entire SED.

In order to construct the $\gamma$-ray energy spectrum for M87 using our model for a jet of relativistic protons emanating from a two-fluid accretion disc, we must vary the three free parameters $\xi$, $E_c$, and $E_{p,\rm max}$ with the goal of obtaining acceptable fits to the TeV $\gamma$-ray spectrum for either the 2004, 2005, or 2010 flare data. The multi-wavelength observations for a given flare are compared with the theoretical spectrum obtained by combining our computation of the TeV emission with the leptonic SSC spectrum for M87 presented by A09. As discussed in Section \ref{secLowEnergyMagDef}, we set the critical energy for cross-field Bohm diffusion into the base of the jet using $E_c=0.624$ TeV. The maximum proton energy $E_{\rm p, max}$ appearing in equations~(\ref{eqPhiHighv2}) and (\ref{eqPhiLow1v2}) determines the slope of the theoretical TeV spectrum, and therefore this parameter varies depending on the data set analyzed. Likewise, the similarity parameter $\xi$ determines the efficiency of the $\gamma$-ray production process, and therefore its value is different for each observed flare.

In Figures~\ref{figCh5C}a, \ref{figCh5C}b, and \ref{figCh5C}c, we plot the multi-wavelength spectra computed using our model and compare it with the observational data for the M87 flares observed in 2010, 2005, and 2004, respectively. The associated parameter values are listed in Tables~\ref{table:tableNew1}, \ref{table:tableNew2}, and \ref{table:tableNew3} for the 2010 {\it VERITAS}, 2005 {\it HESS}, and 2004 {\it HESS} flares, respectively. We note that the fits to the data are reasonably good across the entire multi-wavelength range. For a given value of $\xi$, a range of values exists for the jet half-angle, $\theta$, the cloud radius, $L_0$, and the cloud's proton number density, $n_p$ (see equations~\ref{eqXiMod} and \ref{eqPsiMod}). Once the half-angle $\theta$ and the cloud radius $L_0$ has been specified, we can use equation~(\ref{eqXiMod}) to compute the proton number density $n_p$ for the target cloud or stellar atmosphere. Various authors have adopted different values for the jet half-angle $\theta$. For example, B12 set $\theta=2^{\circ}$, A09 set $\theta=10^{\circ}$, and Walker et al. (2018) set $\theta=17^{\circ}$. We consider these three values of $\theta$ to be a reasonable representation of the expected range for this parameter for M87. We also select three representative values for the variability timescale, $\Delta t$, namely $\Delta t = 5,\,6,\,7$ days, which approximates the observed range for the three flares studied here.

The distance between the black hole and the cloud, $R_c$, is computed using equation~(\ref{eqRc3}), the altitude of the cloud above the disc plane, $z_c$, is computed using equation~(\ref{eqZcRc}), and the jet radius, $r_j$, is computed using equation~(\ref{eqRc1}). Note that the vertical height of the cloud  above the disc, $z_c$, remains greater than the disc half-thickness at the shock location, $H_*$, and the jet radius, $r_j$, exceeds the cloud radius, $L_0$, in satisfaction of equations~(\ref{eqZcHCondition}) and (\ref{eqrjL0Condition}), respectively.

Each data set requires a unique value for the similarity parameter $\xi$, with $\xi=6.21\times10^{25}\,{\rm cm}^{-2},\,1.56\times10^{25}\,{\rm cm}^{-2},\,3.11\times10^{24}\,{\rm cm}^{-2}$ for the 2010, 2005, and 2004 flares, respectively. In each case, the jet half-angle, $\theta$, can have any of the three values considered here, $\theta=2^{\circ},\,10^{\circ},\,17^{\circ}$. Since the jet properties are assumed to remain constant over very long timescales, it follows that the value of $\theta$ should be held constant for all of the flares considered here. This indicates that the different properties of the three TeV flares observed from M87 using {\it VERITAS} and {\it HESS} resulted from differences in the properties of the target cloud or stellar atmosphere, and not from a variation in the jet dynamics. The clouds hit by the M87 jet had different column densities for each data set, with the highest column density observed in 2010, and with smaller values in 2004 and 2005. This suggests that the 2010 flare observed by {\it VERITAS} was due to a rather dense (or large) cloud passing through the jet.

Out of the three values for the jet opening half-angle $\theta$ presented in the tables, we argue that the results obtained for $\theta=10^{\circ}$, adopted by A09, are the most physically reasonable. Setting $\theta=17^{\circ}$ results in a cloud height $z_c$ about equal to the disc half-thickness $H_*$, which is probably too close to the black hole for the model developed here. Furthermore, models with $\theta=2^{\circ}$ are likely to be unphysical due to geometrical restrictions that rule out jets with $\theta<3^{\circ}$ (Biretta et al. 1999). On the other hand, for the value $\theta=10^{\circ}$, all of the restrictions given in equations~(\ref{eqZcHCondition}) for the altitude of the cloud above the disc plane, and (\ref{eqrjL0Condition}) for the radius of the jet with the respect to the cloud radius, are satisfied. We note that our values for the cloud proton number density and cloud radius are similar to those obtained by B12 in their analysis of the 2010 flare.

\subsection{$\gamma$-ray Attenuation}
An important issue for the observation of TeV $\gamma$-rays is the possible attenuation of the high-energy radiation inside the target cloud. The primary attenuation mechanism for TeV emission is the production of electron-positron pairs via collisions between the $\gamma$-ray and either a proton, an electron, or another photon. We discuss these three possibilities in detail below. The expressions for the cross sections describing the various processes are taken from Svensson (1982, hereafter S82).

A useful quantity for comparison purposes is the Thomson optical thickness of the target cloud or red giant atmosphere, computed using
\begin{equation}
\tauT = \sigmaT n_eL_0 \ ,
\label{eqTauThom}
\end{equation}
where $\sigmaT$ denotes the Thomson cross section, $n_e$ is the electron number density, and $L_0$ is the radius of the cloud. In a fully-ionized hydrogen cloud, $n_e=n_p$, where $n_p$ is the proton number density. Since the Thomson cross section $\sigmaT \sim10^{-24}\,{\rm cm}^2$, it follows from the values of $n_p$ and $L_0$ listed in Tables~\ref{table:tableNew1}, \ref{table:tableNew2}, and \ref{table:tableNew3} that the Thomson optical depth $\tauT\sim0.01-1.0$. The corresponding optical thickness for each pair production process, $\tau_{ij}$, is computed using
\begin{equation}
\tau_{ij} = \sigma_{ij} n_eL_0 \ ,
\label{eqTauIJ}
\end{equation}
where $\sigma_{ij}$ represents the associated cross section for the process. Comparing equations~(\ref{eqTauThom}) and (\ref{eqTauIJ}), we see that the relative optical depth for pair production, $\tau_{ij}/\tauT$, is equal to the cross section ratio, $\sigma_{ij}/\sigmaT$, and therefore it is sufficient for our purposes to evaluate this ratio for each process of interest here.

\subsubsection{Photon-electron ($\gamma e^{\pm}\to e^{\pm}e^+e^-$) pair production}
The photon-electron (or photon-positron) pair production cross section is described by equation~(32) from S82,
\begin{equation}
\sigma_{\gamma e}(y)=\frac{3\alpha}{8\uppi}\,\sigmaT\left(\frac{28}{9}\ln2y-\frac{218}{27}\right)\qquad (y\gg1) \ ,
\label{eqSigGamE}
\end{equation}
where $\alpha=1/137$ is the fine-structure constant, and $y=E/(m_ec^2)$ is the dimensionless photon energy. For photon energy $E\sim1\,$TeV, we obtain $y\sim10^6$ and therefore $\sigma_{\gamma e}/\sigmaT\sim0.03$. Since $\tauT\lesssim1$, the resulting optical depth for photon-electron pair production inside the cloud is $\tau_{\gamma e}\lesssim10^{-2}$, which suggests that attenuation of the $\gamma$-ray spectrum due to photon-electron pair production is negligible.

\subsubsection{Photon-proton ($\gamma p\to pe^+e^-$) pair production}
The photon-proton cross section is described by equation~(41) from S82,
\begin{equation}
\sigma_{\gamma p}(y)=\frac{3\alpha}{8\uppi}\,\sigmaT\left(\frac{28}{9}\ln2y-\frac{218}{27}\right)\qquad (y\gg1) \ ,
\end{equation}
which is identical to equation~(\ref{eqSigGamE}). Hence, we can immediately conclude that attenuation due to photon-proton pair production inside the cloud is negligible.

\subsubsection{Photon-photon ($\gamma\gamma\to e^+e^-$) pair production}
The photon-photon cross section is described by equation~(24) from S82,
\begin{equation}
\sigma_{\gamma\gamma}(\yCM)=\frac{3}{8\,\yCM^2}\,\sigmaT\left(2\ln2\yCM-1\right)\qquad (\yCM\gg1) \ ,
\end{equation}
where $\yCM$ is defined as the dimensionless photon energy in the CM frame. For photon energy $E\sim1\,$MeV, we obtain $\yCM\sim2$ and therefore $\sigma_{\gamma \gamma}/\sigmaT\sim0.2$, and consequently the corresponding optical thickness is $\tau_{\gamma \gamma}\lesssim10^{-1}$. Since the cross section $\sigma_{\gamma\gamma}$ is a monotonically decreasing function of $\yCM$, it follows that $\tau_{\gamma \gamma}\ll1$ for all $\gamma$-ray energies of interest here. Hence attenuation of the $\gamma$-ray spectrum due to photon-photon pair production is also negligible.

\section{CONCLUSION}
\label{secConclusion}

We have developed a new self-consistent model for the generation of the observed TeV emission from M87 via collisions between a jet of relativistic protons and a cloud or stellar atmosphere located within $\sim 0.01-0.1\,$pc from the central black hole. The model is able to reproduce the complete multi-wavelength SED for the TeV flares observed in 2004, 2005, and 2010. In Paper~1, we analyzed in detail the structure of two-fluid ADAF accretion discs, in which the structure of the flow is influenced by the back-reaction of the relativistic particles on the thermal gas. The inclusion of the dynamical effect of the particle pressure leads to the formation of a characteristic deceleration precursor, that smooths and weakens the shock discontinuity, in a manner similar to that seen in studies of cosmic-ray modified shocks (Axford et al. 1977; Becker \& Kazanas 2001). Our focus in this paper is on the implications of the two-fluid disc model for the formation of a jet of relativistic protons, which can generate secondary TeV $\gamma$-ray emission via neutral pion decay when the jet encounters a cloud or stellar atmosphere. We explore the implications of the disc and the outflows for the production of TeV $\gamma$-radiation, resulting from collisions between the jet of relativistic protons and a cloud or stellar atmosphere located within one parsec from the central black hole.

We applied the model to the interpretation of a series of high-energy flares observed from M87 by {\it VERITAS} and {\it HESS} in 2004, 2005, and 2010. The scenario we consider here is based on the work of Barkov et al. (2012), who also analyzed the production of TeV $\gamma$-rays due to collisions between a proton jet and a cloud. However, our model provides a unified explanation for the observations, since it includes a physical mechanism for the formation of the proton jet, via particle acceleration occurring around the standing shock in the accretion disc. A rigorous mathematical method was employed to obtain the analytical solution for the Green's function describing the relativistic proton distribution in the disc, which is given by equation~(\ref{eqGreensExpansionV1}). The self-consistency of the model was confirmed via a comparison between two methods for computing the energy density of the relativistic protons in the disc, denoted by $U_{\rm rel}(r)$. One method employs numerical integration of the governing differential equation~(\ref{eqVertIntTransUr}) for $U_{\rm rel}(r)$, and the other employs term-by-term integration of the series expansion for the relativistic proton Green's function (equation~\ref{eqGreensExpansionV1}), which yields equation~\ref{eq51LB07}. The excellent agreement between the two sets of results for $U_{\rm rel}(r)$, plotted in Figure~\ref{figMeanEnergyDyn}, confirms the validity of our solution method.

The particle acceleration model developed here is based on the presence of a standing shock located near the centrifugal barrier in ADAF discs. The possible existence of such shocks was first explored in the context of steady-state models by Chakrabarti (1989), Chakrabarti \& Molteni (1993), and Lu \& Yuan (1997). The question was further investigated by Hawley, Smarr \& Wilson (1984a,b), who demonstrated the existence of standing shocks in tenuous discs using relativistic 2D simulations. Similar results have also been obtained recently by Dihingia et al. (2019), Kumar \& Gu (2019a,b), and Sarkar \& Chattopadhyay (2019). The stability of discs with standing shocks and outflows was questioned by Okuda \& D. Molteni (2012) in their study of accretion onto Sgr~A*, but a subsequent study by Le et al. (2016) established the stability of ADAF discs with standing shocks over a range of values for the viscosity and angular momentum of the accreting gas.

We have demonstrated that the hadronic TeV emission model developed here can be combined with the one-zone leptonic SSC model of Finke et al. (2008) to successfully reproduce the multi-wavelength SED for each of the flares observed from M87 in 2004, 2005, and 2010 (see Figures~\ref{figCh5C}a, \ref{figCh5C}b, and \ref{figCh5C}c). We argue that a superposition of the hadronic and leptonic emission components is reasonable since the two radiation components are emitted by distinct populations of particles that need not be cospatial. The results plotted in Figures~\ref{figCh5C}a, \ref{figCh5C}b, and \ref{figCh5C}c represent the first time the TeV flares have been directly connected with physical processes operating in the accretion disc. We find that the properties of the flares observed using {\it VERITAS} in 2010 and {\it HESS} in 2004 and 2005 can be explained in terms of a collision between a jet of relativistic protons and a cloud or stellar atmosphere with proton number density $n_p \sim 10^9 -10^{10}\,{\rm cm}^{-3}$, and radius $L_0 \sim 10^{13}-10^{14}\,$cm, in Keplerian motion $\sim 10^{16}\,$cm from the central black hole. The mean isotropic Lorentz factor of the protons striking the cloud is $\langle\gamma\rangle_{\rm jet} \sim 10^3$ for each of the flares observed in 2004, 2005, and 2010, suggesting that the dynamics of the M87 jet did not change, but instead the jet collided with clouds of differing properties to produce the three distinct flare spectra.

Recent observational studies indicate that the M87 jet has a variability timescale in the X-rays of $\sim3\,$weeks (Harris et al. 2009). Furthermore, VLBI studies of the observed radio knots yield a similar timescale, which is also consistent with the estimated synchrotron cooling timescale for the radiating electrons (Hada et al. 2012). Assuming that the mass of the central black hole in M87 is given by $M=6.5\times 10^9\,\msun$ (Akiyama et al. 2019), the light crossing time for one gravitational radius is $\sim0.37\,$days, which is the lower bound for the disc to relax to a new steady state if any of the model parameters or boundary conditions were varied. In our application to M87, we find that the flow velocity upstream from the shock is $\sim0.14c$. Hence, we estimate that any variations in the model parameters, or the accretion rate, or the shock location (due to shock oscillation) would lead to relaxation on a timescale of $\sim2.6\,$days. We therefore conclude that the model is able to relax to a new steady-state configuration on a timescale that is much shorter than the observed variability timescale. This implies that a steady-state model of the sort investigated here can be used to interpret the data for a source with the variability behavior exhibited by M87.

We note that the theoretical spectrum plotted in Figure~\ref{figCh5C}c using our model is quite similar to the that displayed in Figure~2 from Fraija \& Marinelli (2016) for the 2004 flare observed by {\it HESS}, which is not surprising since they treated the closely related process of p$\gamma$ pion production, rather than the pp process considered here. Following the transition from LB05 to LB07, we plan to study the effect of viscosity on the structure of the disc and the formation of the standing shock and the associated relativistic outflow. We expect that the inclusion of viscosity will not significantly alter the conclusions reached in this work, since significant particle acceleration will occur regardless of the level of viscosity, provided that a shock is present. In particular, we will reexamine the question of whether both shocked and smooth flow solutions are possible when particle diffusion and viscosity are both included. It is also interesting to note that the jets of relativistic protons considered here may also be efficient sources of cosmological neutrinos, although we have not made any estimates regarding this possibility yet (e.g. Righi et al. 2018; Zhang et al. 2016; Reville \& Bell 2014). We conclude that our coupled, self-consistent theory for the disc structure and the associated particle acceleration provides for the first time a completely self-consistent explanation for the outflows and the high-energy $\gamma$-ray emission observed in radio-loud AGNs.

\section*{Acknowledgements}

The authors are grateful to the anonymous referee whose comments and suggestions helped to significantly improve the manuscript.

\appendix

\section[Determining the Eigenfunction Jump Condition]{Eigenfunction Shock Jump Condition}
\label{AppendixEigenJump}

The global solution for the eigenfunction $Y_n(r)$ must satisfy the continuity and derivative jump conditions associated with the presence of the shock/source at radius $r=r_*$. In order to obtain these conditions, we must integrate the transport equation with respect to radius in the vicinity of the shock. Beginning with equation~(\ref{eqDYnDr2}), we have
\begin{equation}
-H\vel\frac{dY_n}{dr} = \frac{\lambda_n}{3r}\frac{d}{dr}(rH\vel)Y_n + \frac{1}{r}\frac{d}{dr}\left(rH\kappa\frac{dY_n}{dr}\right)-A_0 c H_*\delta(r-r_*)Y_n \ .
\label{eqA1}
\end{equation}
Multiplying both sides by $r$ and integrating with respect to radius in the vicinity of the shock yields
\begin{equation}
\lim_{\epsilon\to 0} \int_{r_*-\epsilon}^{r_*+\epsilon} - H\vel r\frac{dY_n}{dr}dr
= \lim_{\epsilon\to 0} \frac{\lambda_n}{3}\int_{r_*-\epsilon}^{r_*+\epsilon}\frac{d}{dr}(rH\vel)Y_n dr
+ \int_{r_*-\epsilon}^{r_*+\epsilon}\frac{d}{dr}\left(rH\kappa\frac{dY_n}{dr}\right)dr - A_0 c H_*\int_{r_*-\epsilon}^{r_*+\epsilon} r\delta(r-r_*)Y_n dr \ .
\label{eqA2}
\end{equation}
Since the left-hand side of this expression contains no singular factors, it vanishes in the limit $\epsilon \to 0$, and we therefore obtain
\begin{equation}
0 = \lim_{\epsilon\to 0} \frac{\lambda_n}{3} \int_{r_*-\epsilon}^{r_*+\epsilon}\frac{d}{dr}(rH\vel)Y_n dr
- \Delta\left[rH\kappa\frac{dY_n}{dr}\right] - A_0 c H_* r_* Y_n(r_*) \ ,
\label{eqA3}
\end{equation}
where $\Delta$ denotes the difference between post-shock and pre-shock quantities (cf. equation~\ref{eqDeltaM2}),
\begin{equation}
\Delta[f] \equiv \lim_{\delta\to 0} f(r_*-\delta) - f(r_*+\delta) = f_+ - f_- \ .
\label{eqA4}
\end{equation}
Applying integration by parts to the first term on the right-hand side of equation~(\ref{eqA3}) yields
\begin{equation}
0 = - \frac{\lambda_n}{3}\Delta\left[rH\vel Y_n\right]
- \frac{\lambda_n}{3} \lim_{\epsilon\to 0} \int_{r_*-\epsilon}^{r_*+\epsilon} rH\vel\frac{dY_n}{dr}dr
- \Delta\left[rH\kappa\frac{dY_n}{dr}\right] - A_0 c H_* r_* Y_n(r_*) \ .
\label{eqA5}
\end{equation}
The second term on the right-hand side of this expression contains no singularities, and therefore it vanishes in the limit $\epsilon \to 0$. Hence equation~(\ref{eqA5}) reduces to
\begin{equation}
\Delta\left[\frac{\lambda_n}{3}H\vel Y_n+H\kappa\frac{dY_n}{dr}\right] = -A_0cH_*Y_n(r_*) \ .
\label{eqA6}
\end{equation}
Equation~(\ref{eqA6}) gives the derivative jump in the context of the two-fluid model. We note that the function $Y_n(r)$ itself must be continuous at $r=r_*$ in order to avoid an infinite diffusive flux there, and this yields the continuity condition
\begin{equation}
\Delta\left[Y_n\right] = 0 \ .
\label{eqA7}
\end{equation}
Equations~(\ref{eqA6}) and (\ref{eqA7}) establish that $Y_n(r)$ must be continuous at the shock location, and its derivative must display a jump there.

\section[Asymptotic Relations for the Eigenfunction]{Asymptotic Eigenfunctions Relations}
\label{AppendixEigenAsymptotic}

In this section we derive the physical boundary conditions satisfied by the spatial eigenfunctions $Y_n(r)$. The boundary conditions are combined with the jump conditions in order to determine the global eigenfunctions and the associated eigenvalues $\lambda_n$. We begin with equation~(\ref{eqDYnDr}), which states that
\begin{equation}
\frac{d^2Y_n}{dr^2} + \left[\frac{\rs }{\kappa_0(r-\rs)^2} + \frac{d\ln(rH\vel )}{dr} + \frac{2}{r-\rs}\right]\frac{dY_n}{dr} + \frac{\lambda_n\rs}{3\kappa_0(r-\rs )^2}\frac{d\ln(rH\vel )}{dr}\,Y_n = 0 \ .
\label{eqEigenB1}
\end{equation}
We will analyze this expression in order to obtain boundary conditions for $Y_n(r)$ applicable near the event horizon ($r \to \rs$) and also at large radii ($r \to \infty$).

\subsection{Near the Horizon ($r\to\rs$)}

In Paper~1, we established that near the event horizon, the radial velocity $\vel$ approaches free-fall velocity, $\vel^2_{\rm ff}\equiv 2GM/(r-\rs)$, so that
\begin{equation}
\vel \propto(r-\rs )^{-1/2} \ , \qquad r\to\rs \ .
\label{eqEigenB2}
\end{equation}
We also demonstrated that with the inclusion of relativistic particle pressure, the asymptotic variation of the disc half-thickness $H$ near the horizon is given by (see equation~C7 from Paper~1)
\begin{equation}
H\propto(r-\rs)^{(\gammag + 3)/[2(\gammag + 1)]} \ , \qquad r\to\rs \ .
\label{eqEigenB3}
\end{equation}
Applying equations~(\ref{eqEigenB2}) and (\ref{eqEigenB3}) to the logarithmic term in equation~(\ref{eqEigenB1}) yields
\begin{equation}
\frac{d\ln(rH\vel)}{dr} \approx [(\gammag + 1)(r-\rs)]^{-1} \ , \qquad r\to\rs \ , 
\label{eqEigenB4}
\end{equation}
which reduces equation~(\ref{eqEigenB1}) to
\begin{equation}
\frac{d^2Y_n}{dr^2} + \left[\frac{\rs }{\kappa_0(r-\rs )^2} + \left(\frac{1}{\gammag + 1} + 2\right) \frac{1}{r-\rs}\right]\frac{dY_n}{dr}
+ \frac{\lambda_n\rs }{3\kappa_0(r-\rs )^3(\gammag + 1)}\,Y_n = 0 \ , \qquad r\to\rs \ .
\label{eqEigenB5}
\end{equation}

One can see that equation~(\ref{eqEigenB5}) is equivalent to equation~(C14) from Paper~1, if we make the identification $\lambda\to n+1$. Hence we can apply the same Frobenius approach utilized in Appendix~C from Paper~1 to immediately conclude that near the event horizon, the asymptotic variation of the spatial eigenfunction $Y_n(r)$ is given by
\begin{equation}
Y_n(r)\propto(r-\rs)^{-\lambda_n/(3\gamma_{\rm th} + 3)} \ , \qquad r\to\rs \ ,
\label{eqEigenB7}
\end{equation}
which is the same asymptotic behavior obtained by LB07 in the context of their one-fluid model. In Figure~\ref{figGginGgout1}a, we plot a sample comparison between the fundamental numerical solutions $G^{\rm in}_n(r)$ and the corresponding inner asymptotic functions $g_n^{\rm in}(r)$ for $n=1,2,3$. Note that the two functions agree closely in the limit $r\to\rs$, as expected. This validates our utilization of the inner asymptotic form (equation~\ref{eqGin}) in setting the inner boundary condition for the spatial eigenfunctions $Y_n$.

\subsection{Towards Infinity ($r\to\infty$)}

\begin{figure}
\centering
\includegraphics[width=0.8\textwidth]{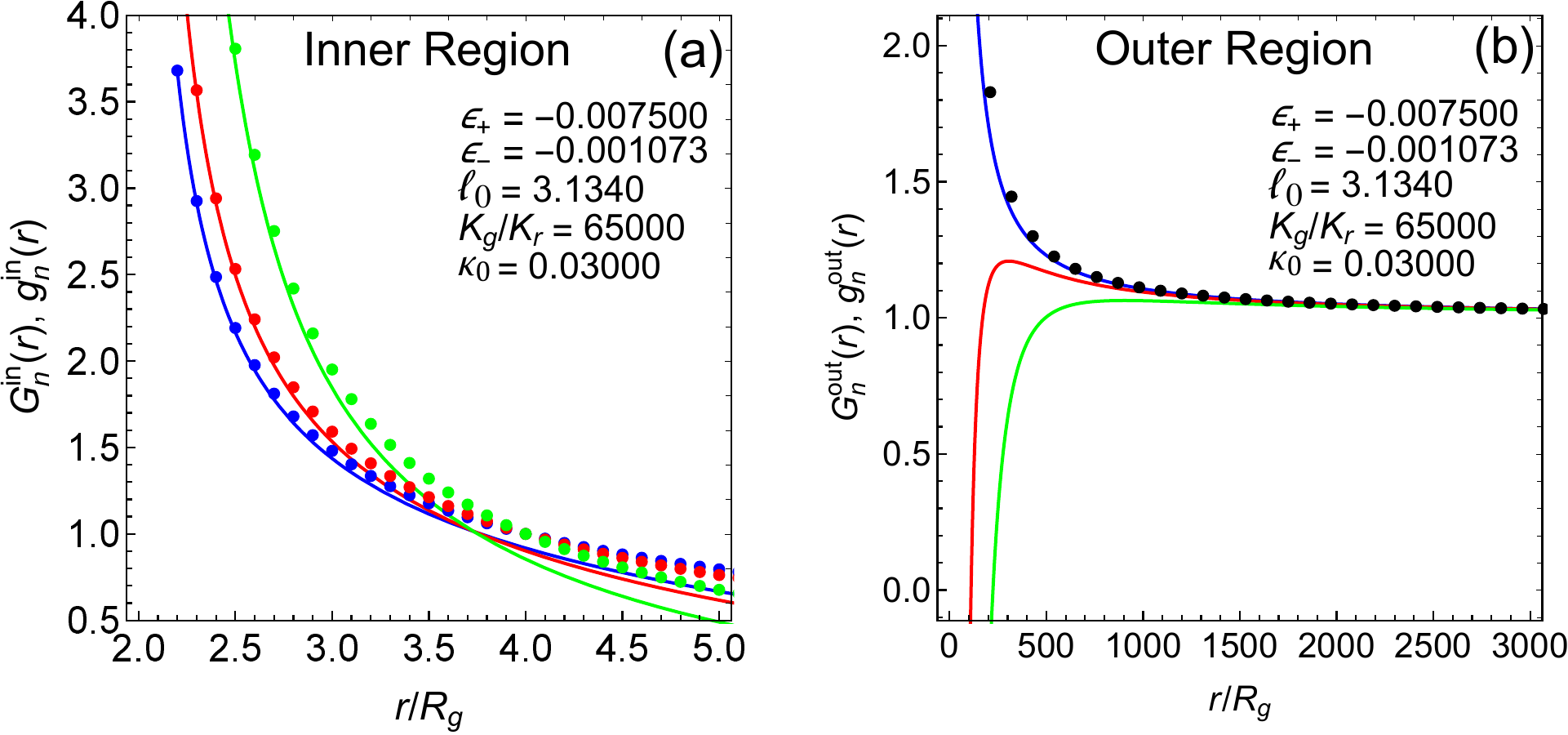}
\caption{Fundamental solutions obtained in Model~C for a) $G^{\rm in}_n(r)$ (equation~\ref{eqGin}) and b) $G^{\rm out}_n(r)$ (equation~\ref{eqGout}). The blue, red, and green values correspond to $n=1, 2, 3$, respectively. The solutions are compared with the corresponding asymptotic forms $g^{\rm in}_n(r)$ and $g^{\rm out}_n(r)$ (filled circles).}
\label{figGginGgout1}
\end{figure}

In Paper~1, we established that particle transport is dominated by outward-bound spatial diffusion as $r\to\infty$. This leads to the determination of the asymptotic behavior of the inflow velocity $\vel$ (see equation C9 from Paper~1), given by
\begin{equation}
\vel \propto r^{-5/2} \ , \qquad r\to\infty \ ,
\end{equation}
as well as the asymptotic behavior of the disc half-thickness $H$ (see equation C10 from Paper~1), given by
\begin{equation}
H\propto r^{3/2} \ , \qquad r\to\infty \ .
\end{equation}
Application of these two relations to the logarithmic term in equation~(\ref{eqEigenB1}) yields
\begin{equation}
\frac{d\ln(rH\vel)}{dr} = 0 \ , \qquad r\to\infty \ ,
\label{eqEigenB10}
\end{equation}
which reduces equation~(\ref{eqEigenB1}) to
\begin{equation}
\frac{d^2Y_n}{dr^2} + \left[\frac{\rs}{\kappa_0 r^2} + \frac{2}{r}\right]\frac{dY_n}{dr} = 0 \ , \qquad r\to\infty \ .
\label{eqEigenB11}
\end{equation}
In the asymptotic regime $r\to\infty$, the dominant term inside the square brackets is the one proportional to $1/r$, and therefore we obtain
\begin{equation}
\frac{d^2 Y_n}{dr^2} = -\frac{2}{r}\frac{dY_n}{dr} \ , \qquad r\to\infty \ .
\label{eqEigenH29}
\end{equation}
Upon integration, we obtain the asymptotic form
\begin{equation}
Y_n = \frac{C_1}{r} + C_0 \ , \qquad r\to\infty \ ,
\label{eqEigenH30}
\end{equation}
where $C_0$ and $C_1$ are constants of integration. In Figure~\ref{figGginGgout1}b, we plot a sample comparison between the fundamental numerical solution $G^{\rm out}_n(r)$ and the corresponding outer asymptotic function $g_n^{\rm out}(r)$ for $n=1,2,3$. We observe that the two functions agree closely in the limit $r\to\infty$, as expected. This validates our utilization of the asymptotic form in setting the outer asymptotic form (equation~\ref{eqGout1}) for the spatial eigenfunctions $Y_n(r)$.

\section{Orthogonality of the Spatial Eigenfunctions}
\label{AppendixOrtho}

We can establish the orthogonality of the spatial eigenfunctions $Y_n(r)$ by writing equation~(\ref{eqDYnDr}) in the equivalent Sturm-Liouville form,
\begin{equation}
\frac{d}{dr}\left[S(r)\frac{dY_n}{dr}\right] + \lambda_n\omega(r)Y_n(r) = 0 \ .
\label{eqSturmLiouville}
\end{equation}
In our application, the function $S(r)$ is computed using
\begin{equation}
S(r) \equiv \frac{rH\kappa}{r_* H_* \kappa_*} \exp\left\{\frac{1}{\kappa_0}\left[\left(\frac{r_*}{\rs}
- 1\right)^{-1} - \left(\frac{r}{\rs}-1\right)^{-1}\right]\right\} \ ,
\label{eqSFunction}
\end{equation}
and the weight function $\omega(r)$ is defined by
\begin{equation}  
\omega(r)\equiv\frac{\vel S}{3\kappa}\frac{d\ln(rH\vel)}{dr} \ .
\label{eqWeightFunction}
\end{equation}

Let us suppose that $\lambda_n$ and $\lambda_m$ denote two distinct eigenvalues $(\lambda_n\ne\lambda_m)$ with associated spatial eigenfunctions $Y_n(r)$ and $Y_m(r)$, respectively. Since $Y_n$ and $Y_m$ each satisfy equation~(\ref{eqSturmLiouville}) for their respective eigenvalues, we can write
\begin{equation}
Y_n(r)\left\{\frac{d}{dr}\left[S(r)\frac{dY_m}{dr}\right] + \lambda_m\omega(r)Y_m(r)\right\} = 0 \ ,
\label{eqOrtho2}
\end{equation}
and
\begin{equation}
Y_m(r)\left\{\frac{d}{dr}\left[S(r)\frac{dY_n}{dr}\right] + \lambda_n\omega(r)Y_n(r)\right\} = 0 \ .
\label{eqOrtho3}
\end{equation} 
Subtracting equation~(\ref{eqOrtho3}) from equation~(\ref{eqOrtho2}) yields
\begin{equation}
Y_n(r)\frac{d}{dr}\left[S(r)\frac{dY_m}{dr}\right] - Y_m(r)\frac{d}{dr}\left[S(r)\frac{dY_n}{dr}\right]
= (\lambda_n - \lambda_m)\omega(r)Y_n(r)Y_m(r) \ .
\label{eqOrtho4}
\end{equation} 
We can integrate equation~(\ref{eqOrtho4}) by parts from $r=\rs$ to $r=\infty$ to obtain, upon simplification,
\begin{equation}
S(r)\left[Y_n(r)\frac{dY_m}{dr} - Y_m(r)\frac{dY_n}{dr}\right]_{\rs}^{\infty}
= (\lambda_n - \lambda_m)\int_{\rs}^{\infty}\omega(r)Y_n(r)Y_m(r)dr \ .
\label{eqOrtho6}
\end{equation}

The asymptotic behaviors of the inner and outer fundamental solutions, $G_n^{\rm in}(r)$ and $G_n^{\rm out}(r)$, respectively, are stated in equations (\ref{eqGin}) and (\ref{eqGout1}) for the limits $r\to\rs$ and $r\to\infty$, respectively. By virtue of equation~(\ref{eqYnCases}), the spatial eigenfunctions $Y_n(r)$ obey the same set of boundary conditions. Based on these conditions, we conclude that the left-hand side of equation~(\ref{eqOrtho6}) vanishes, leaving
\begin{equation}
\int_{\rs}^{\infty}\omega(r)Y_n(r)Y_m(r)dr = 0, \quad m \ne n \ .
\label{eqOrtho7}
\end{equation} 
This result establishes that $Y_m$ and $Y_n$ are orthogonal eigenfunctions relative to the weight function $\omega(r)$ defined in equation~(\ref{eqWeightFunction}).
Note that the weight function $\omega(r)$ displays a $\delta$-function behavior at $r=r_*$ due to the variation of the derivative $\vel'(r)$ in the vicinity of the shock. In this region, we can combine equations (\ref{eqDvDrDelta}), (\ref{eqSFunction}), and (\ref{eqWeightFunction}) to show that
\begin{equation}
\omega(r)\to\frac{1}{3\kappa_* H_*}(H_- \vel_- - H_+\vel_+)\delta(r-r_*) \ , \quad r\to r_* \ ,
\label{eqWeightFunction2}
\end{equation}
which is a generalization of the weight function given by equation~(40) from LB07, applicable for the two-fluid model considered here.

\section{Expansion Coefficients}
\label{secAppExp}

The exact solution for the Green's function, $\greens(\Eproton,r)$, is given by the series expansion (cf. equation~\ref{eqGreensExpansionV1})
\begin{equation}
\greens(\Eproton,r) = \sum^{N_{\rm max}}_{n=1}b_nY_n(r)\left(\frac{\Eproton}{E_0}\right)^{-\lambda_n} \ , \quad \Eproton \geq E_0 \ ,
\label{eqD1}
\end{equation}
where $Y_n(r)$ denotes the set of spatial eigenfunctions. In order to evaluate the Green's function using equation~(\ref{eqD1}), we require knowledge of the expansion coefficients, $b_n$. These coefficients can be computed by exploiting the orthogonality of the spatial eigenfunctions as follows. We begin by noting that for proton energy $E_p=E_0$, equation~(\ref{eqD1}) reduces to
\begin{equation} 
\greens(\Eproton,r) = \sum_{m=1}^{N_{\rm max}}b_mY_m(r) \ .
\label{eqD2}
\end{equation}
Multiplying both sides of equation~(\ref{eqD2}) by the product $Y_n(r)\omega(r)$ and integrating with respect to $r$ from $r=\rs$ to $r=\infty$ yields
\begin{equation}  
\int_{\rs}^{\infty}\greens(E_0,r)Y_n(r)\omega(r)dr = \sum_{m=1}^{N_{\rm max}}b_m\int_{\rs}^{\infty}Y_m(r)Y_n(r)\omega(r)dr \ .
\label{eqD3}
\end{equation}
Based on the orthogonality of the spatial eigenfunctions (equation~\ref{eqOrtho7}), we observe that only the $n=m$ term on the right-hand side of equation~(\ref{eqD3}) survives, leaving
\begin{equation}
\int_{\rs}^{\infty}\greens(E_0,r)Y_n(r)\omega(r)dr = b_n\int_{\rs}^{\infty}Y_n^2(r)\omega(r)dr \ .
\label{eqD4}
\end{equation}
Hence the expansion coefficient $b_n$ can be expressed as
\begin{equation}  
b_n = \frac{\int_{\rs}^{\infty}\greens(E_0,r)Y_n(r)\omega(r)dr}{{\cal I}_n} \ ,
\label{eqD5}
\end{equation}
where the quadratic normalization integral, ${\cal I}_n$, is defined by
\begin{equation}
{\cal I}_n\equiv\int_{\rs}^{\infty} Y_n^2(r)\omega(r)dr \ .
\label{eqD6}
\end{equation}

Completing the calculation of the expansion coefficients, $b_n$, for our two-fluid model requires the evaluation of the distribution function at the source energy, $\greens(E_0,r)$. We can obtain an expression for this quantity by integrating equation~(\ref{eqVertTransport}) with respect to $\Eproton$ in a small range around the injection energy $E_0$, obtaining
\begin{equation}
0 = \frac{1}{3r}\frac{\partial}{\partial r}(r H \vel_r)E_0\greens(E_0,r)
+ \frac{\dot N_0\delta(r-r_*)}{(4\uppi\, E_0)^2 r_*} \ ,
\label{eqD7}
\end{equation}
where we have used the fact that $\greens(E_p,r)=0$ for $E_p<E_0$. Equation~(\ref{eqD7}) clearly indicates that $\greens(E_p,r)=0$ for $r \ne r_*$. The value of $\greens(E_p,r_*)$ can be obtained by integrating equation~(\ref{eqD7}) with respect to $r$ over a small region surrounding the shock location, which yields
\begin{equation}
0 = \frac{1}{3} (H_+ \vel_+ - H_- \vel_-) E_0\greens(E_0,r_*)
+ \frac{\dot N_0}{(4\uppi\, E_0)^2 r_*} \ ,
\label{eqD8}
\end{equation}
Combining relations, we find that
\begin{equation} 
\greens(E_0,r) = \begin{cases}
\frac{3\dot N_0}{(4\uppi\,)^2 E_0^3 r_* (H_-\vel_- - H_+\vel_+)}, & r=r_* \ , \\
0, & r\ne r_* \ .
\label{eqD9}
\end{cases}
\end{equation}
Substituting for $\greens(E_0,r)$ in equation~(\ref{eqD5}) using equation~(\ref{eqD9}) and carrying out the integration, we obtain the final result
\begin{equation}  
b_n = \frac{\dot N_0 Y_n(r_*)}{(4\uppi\,)^2 E_0^3 r_* H_* \kappa_* {\cal I}_n} \ , 
\label{eqD10}
\end{equation}
where we have utilized the $\delta$-function behavior close to the shock for the weight function $\omega(r)$ given by equation~\ref{eqWeightFunction2}. Formally, equation (\ref{eqD10}) is exactly the same as equation~(49) from LB07. However, when we also consider the singular nature of the weight function when computing the normalization integrals ${\cal I}_n$ defined in equation~(\ref{eqD6}), 
\begin{equation}
{\cal I}_n = \lim_{\epsilon\to0}\int_{\rs}^{r_* - \epsilon}\omega(r)Y_n^2(r)dr
+ \int_{r_* + \epsilon}^{\infty}\omega(r)Y_n^2(r)dr
+ \frac{1}{3\kappa_* H_*}(H_- \vel_- - H_+ \vel_+)Y^2_n(r_*) \ ,
\label{eqInFinal}
\end{equation}
we find that ${\cal I}_n$ is different in our two-fluid model, because of the discontinuity of the disc half-thickness $H$ at the shock radius $r_*$.

\bsp	
\label{lastpage}

\end{document}